\documentclass[fleqn,usenatbib]{mnras}

\usepackage{newtxtext,newtxmath}

\usepackage[T1]{fontenc}

\DeclareRobustCommand{\VAN}[3]{#2}
\let\VANthebibliography\thebibliography
\def\thebibliography{\DeclareRobustCommand{\VAN}[3]{##3}\VANthebibliography}

\usepackage{graphicx}	
\usepackage{amsmath}	

\newcommand{\msun}{{\,\rm M_\odot}}

\newcommand{\kms}{\,{\rm km}\,{\rm s}^{-1}}

\def\gsim{ \lower .75ex \hbox{$\sim$} \llap{\raise .27ex \hbox{$>$}} }
\def\lsim{ \lower .75ex \hbox{$\sim$} \llap{\raise .27ex \hbox{$<$}} }

\title[Satellites vs. dark matter models]{Matching the mass function of Milky Way satellites in competing dark matter models}

\author[M.~R.~Lovell]{
Mark R. Lovell$^{1}$\thanks{E-mail: lovell@hi.is} and Jes\'us Zavala$^{1}$
\\
$^{1}$Centre for Astrophysics and Cosmology, Science Institute, University of Iceland, Dunhaga 5, 107 Reykjav\'ik, Iceland}

\date{Accepted 2023 January 18. Received 2023 January 18; in original form 2022 September 26}

\pubyear{2023}

\begin{document}
\label{firstpage}
\pagerange{\pageref{firstpage}--\pageref{lastpage}}
\maketitle

\begin{abstract}
 Any successful model of dark matter must explain the diversity of observed Milky Way (MW) satellite density profiles, from 
 very dense ultrafaints to low-density satellites so large that they could be larger than their inferred dark matter haloes. Predictions for these density profiles are complicated by the limitations of simulation resolution in the stripping of subhaloes by the MW system. We consider cold dark matter (CDM), warm dark matter (WDM, 3.3~keV thermal relic power spectrum), and a self-interacting dark matter model (SIDM) that induces gravothermal collapse in low-mass subhaloes. Using $N$-body simulations combined with a halo stripping algorithm, we find that most CDM and WDM subhaloes of mass $>10^{8}$~$\msun$ are large enough after stripping to fit most satellites; however, the required amount of stripping often requires a stronger tidal field than is available on the subhalo's orbit. The lower concentrations of WDM subhaloes enable more stripping to take place, even on orbits with large pericentres. SIDM cores offer the best fits to massive, low-density satellites at the expense of predicting $>10^{9}$~$\msun$ subhaloes to host low-density satellites with no observed analogue. The agreement of the total number of satellites with observations in CDM and WDM depends strongly on the assumptions made to draw the observational estimates. We conclude that an SIDM model must have a very high velocity-dependent cross-section in order to match all satellites, and that WDM offers a marginally better fit than CDM to the MW satellite mass function.

\end{abstract}

\begin{keywords}
Local Group -- dark matter
\end{keywords}



\section{Introduction}

 One of the key observables for discriminating between competing dark matter models is the shape of the dark matter halo density profile. The cold dark matter model (CDM) generates haloes that follow a Navarro--Frenk--White profile \citep[NFW;][]{NFW_96,NFW_97}, with a cusp of logarithmic slope $-1$ in the halo centre and slope $-3$ in the outskirts. In the self-interacting dark matter model (SIDM), by contrast, the self-interactions scatter dark matter out of the halo centre to form a cored profile \citep{Spergel00,Vogelsberger12}. This core can later be completely erased if the rate of self-interactions is high enough to initiate gravothermal collapse, in which case the dark matter rapidly flows to the halo centre and generates a very steep cusp \citep{Balberg02,Colin02,Koda11,Zavala19a,Correa22}. 

The CDM cusp and the SIDM core scenarios are bridged to a degree by the warm dark matter model (WDM). The thermal motions of WDM particles erase low-mass structures at high redshift, the absence of which delays the onset of structure formation to times at which the Universe is less dense \citep{Lovell12,Lovell19a}. The resulting density profile is still cuspy\footnote{The thermal motions of WDM particles will generate a core via phase-space constraints as described by Liouville's theorem. However, this core has a characteristic size of only a few pc and is therefore unlikely to be relevant for dark matter densities measured in satellite galaxies \citep{Dalcanton01,Shao13,Maccio13,Alvey21}.} but has a lower concentration at fixed halo mass relative to CDM at masses lower than $\sim100$ times the half-mode mass \citep{Bose16a,Maccio19}. Finally, models in which there are interactions between dark matter particles via relativistic species (photons, neutrinos, and dark photons) experience both WDM-style free-streaming and late-time self-interactions, thus exhibiting both lower concentrations and cores simultaneously, as described by the ETHOS dark matter framework \citep{Vogelsberger16}.

The potential for divergence in the predictions of dark matter density profiles in CDM, WDM, and SIDM is strongest at the scale of dwarf galaxies, particularly dwarf spheroidal (dSph) satellites of the Milky Way (MW). These systems exhibit very low baryon fractions, and so the degeneracy of dark matter physics with astrophysical processes such as adiabatic contraction and supernova feedback \citep{Gnedin04} is minimized. It is also at the smallest scales that models have the greatest freedom to diverge from collisionless CDM, which has proven to be a good match to observations on scales larger than massive galaxies \citep{Eisenstein05,Planck16}. This divergence in dSph property predictions between dark matter models is also not restricted to halo profile shapes: some of these models also predict different numbers of MW satellites \citep{Bode01,Schneider2012,Vogelsberger12,Lovell14} and different satellite ages \citep{Lovell12,Maccio19,Lovell20c}. The challenge presented to each of these dark matter models is then twofold: (i) is it possible to fit each observed MW satellite with a subhalo from the dark matter model, and (ii) can we accommodate all satellites simultaneously around the MW in all of the subhaloes massive enough to form a galaxy?  

Measurements of MW satellite galaxy densities over the past decade have variously led the salience of alternatives to the CDM model to sporadically increase and decrease. First, the too-big-to-fail problem, as posed in \citet{BoylanKolchin11,BoylanKolchin12}, is the inference that the most massive CDM subhaloes around MW host haloes were too dense to host any of the observed satellite galaxies. Dark matter-based solutions to this problem included WDM-induced lower halo concentrations \citep{Lovell12,Lovell17a,Lovell17b} and SIDM cores \citep{Vogelsberger12,Rocha13,Zavala13}, plus a wide range of less exotic explanations such as revising the MW host halo mass downwards \citep{Wang12}, the evacuation of baryons by reionization \citep{Sawala16b}, and the creation of cores by impulsive feedback \citep{NEF96,Pontzen_Governato_11}. A more recent challenge has been set by the detection of satellite galaxies that are bright ($M_{V}<-8$), yet low in surface brightness and have low velocity dispersions, particularly Crater~II \citep{Torrealba16} and Antlia~II \citep{Torrealba19}. \citet{Errani22} and \citet{Borukhovetskaya22} have argued that it is not possible to strip CDM haloes to match the observed Crater~II mass and size without disrupting the satellite completely, unless the satellite is especially low-density \citep{Amorisco19}, and it may therefore be necessary to alter the pre-stripping DM profile, either through a baryonic core or through novel dark matter physics. \citet{JiA21} instead presented evidence that Antlia~II is in the process of being disrupted, and that Crater~II may be in a very similar situation.

The reported presence of a core in some satellites \citep{Gilmore2007,Walker11} has motivated the assumption of SIDM, yet the very large velocity dispersions measured in the centres of ultrafaint dSphs cannot be matched by cored profiles \citep{Errani18,Zavala19a,Silverman23}, including those of CDM, and therefore an SIDM model would require gravothermal collapse in low-mass objects in order to fit both dense cusps to high-density dSphs and cores in high-mass objects \citep{Zavala19a,Turner21,Correa22}. Finally, the full census of MW satellites across the sky has yet to be achieved, and alternative methods to estimate the total number of satellites from current limits in sky coverage and observation depth come to significantly different conclusions. \citet{Newton18} predicts of order 120 satellites whereas \citet{Nadler20} instead expects $\sim220$, the latter of which will rule out most deviations from CDM.      

A comprehensive assessment of the viability of each dark matter model requires the accurate modelling of a large range of satellite and host properties, including baryonic contraction of the host halo, the MW disc, the impact of baryons, the stochastic variation of satellite infall histories between viable MW host halo-analogues, and the incompleteness of the MW satellite census. There is also evidence that, in some circumstances, even the highest resolution cosmological simulations are unable to capture accurately the stripping of subhaloes \citep{vdBosch18,Green21,Errani21}. When coupled to the fact that the WDM and SIDM models both contain free parameters -- the effective mass of the WDM particle and the velocity-dependent self-interaction cross-section for SIDM -- such a comprehensive approach becomes prohibitive. 

 These challenges are compounded by uncertainties in  in the properties and evolution of the MW halo, and in the mass and orbit of the Large Magellanic Cloud (LMC). Approaches to assess the statistics of halo assembly include, on the one hand, projects that generate large populations of haloes to explore the possible range of MW satellite population properties, examples of which include ELVIS \citep{GarrisonKimmel14}, COCO \citep{Hellwing16}, and the Caterpillar Project \citep{Griffen16}, and on the other hand constrained simulations that aim to match the MW’s own formation history and local environment, such as CLUES \citep{Libeskind10}, HESTIA \citep{Libeskind20} and SIBELIUS \citep{Sawala22}. It is not guaranteed that a simulation constrained to reproduce the environment at scales $\ge1$~Mpc will also match the exact MW satellite spatial distribution and mass function reliably, given these properties are determined at $\sim100$~kpc scales. It is therefore unclear whether one is more likely to obtain a satellite population that matches the ‘true’ MW satellite function by generating a large number of $\sim10^{12}$~$\msun$ haloes or by analysing a single constrained realization.

In this paper we develop a model to obtain first-order estimates of the impact of dark matter physics on the MW satellite density profiles. We subsequently describe the behaviour of each model and how future studies may offer definitive constraints. We consider three dark matter models: CDM; the WDM model for which the matter power spectrum is well approximated by a thermal relic of mass 3.3~keV; and the vd100 model of SIDM \citep{Zavala19a,Turner21}, which adopts an extreme velocity-dependent cross-section in order to trigger gravothermal collapse in low-mass haloes ($<10^{9}$~$\msun$), whilst retaining low cross-sections in high-mass haloes ($>10^{13}$~$\msun$) as per observational constraints. We use host haloes and satellite haloes drawn from the COCO-CDM simulations to model CDM, and hosts and satellites from the counterpart COCO-WDM simulation (thermal relic particle mass: 3.3~keV) as our WDM simulation. Together these simulations are among the largest volume CDM-WDM counterpart runs performed with the resolution necessary to track ultrafaint dwarfs. For SIDM, we impose cores and gravothermal collapse cusps to the COCO-CDM haloes as inspired from the vd100 simulation of \citet{Zavala19a}, and treat the result as an effective `COCO-SIDM' simulation. We apply a stripping model to each subhalo to ascertain whether it is able to host each observed satellite, and compare the results to the observed (although incomplete) satellite mass function.  

This paper is organized as follows. In Section~\ref{sec:sims} we present the simulations, our halo selections, and the stripping algorithm. We compare the results to individual satellites in Section~\ref{sec:indsats} and to the satellite population in Section~\ref{sec:satpop}. In Section~\ref{sec:satprops} we present estimates for satellite properties. We draw conclusions in Section~\ref{sec:conc}.

\section{Simulations, halo selection, and stripping algorithm}
\label{sec:sims}

\subsection{Simulations}

The COCO-CDM simulation is a zoom simulation of a spherical high-resolution region $\sim28$~Mpc in radius, immersed in a cube of side length 100~Mpc \citep{Hellwing16}. The mass of the high-resolution region simulation particle mass is $1.6\times10^{5}$~$\msun$, and the gravitational softening length $\epsilon=0.33$~kpc. The cosmological parameters are consistent with the WMAP7 cosmology \citep{wmap11}: Hubble parameter $h=0.704$, matter density $\Omega_{0}=0.272$, dark energy density, $\Omega_{\Lambda}=0.728$; spectral index, $n_\rmn{s}=0.967$; and power spectrum normalization $\sigma_{8} = 0.81$. We anticipate that the application of the \citet{Planck16} cosmological parameter values will affect our results at the $<5$~per~cent level \citep{Lovell18b}. It was performed with the {\sc p-gadget3} galaxy formation code, which is based on the publicly available {\sc gadget2} code \citep{Springel05}. Haloes are identified with the friends-of-friends (FoF) algorithm, and are subsequently decomposed into a smooth host halo plus gravitationally bound subhaloes using the {\sc subfind} code \citep{Springel01}. The minimum number of particles required to identify a subhalo is 20. 

\begin{figure*}
    \centering
    \includegraphics[scale=0.34]{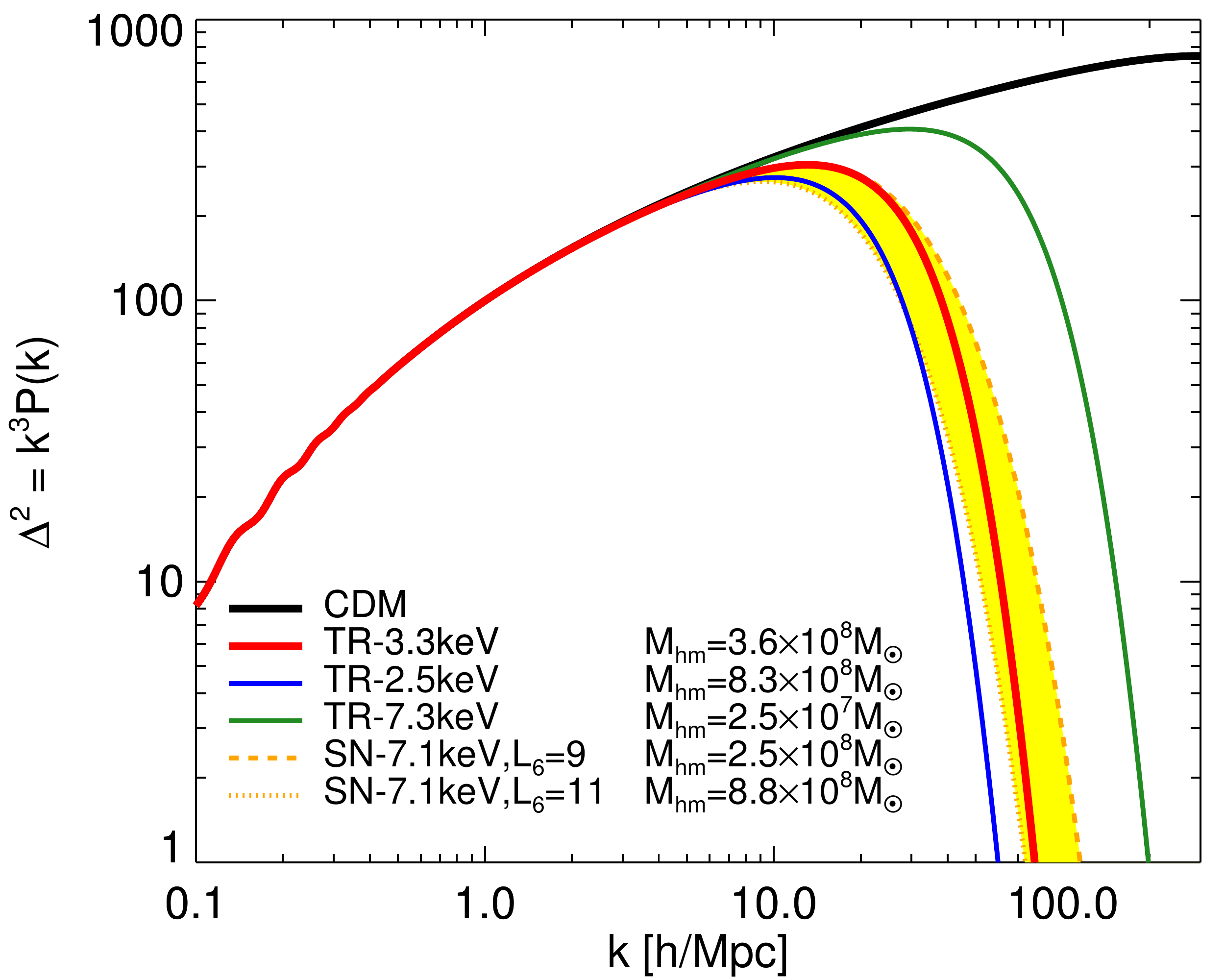}
       \includegraphics[scale=0.34]{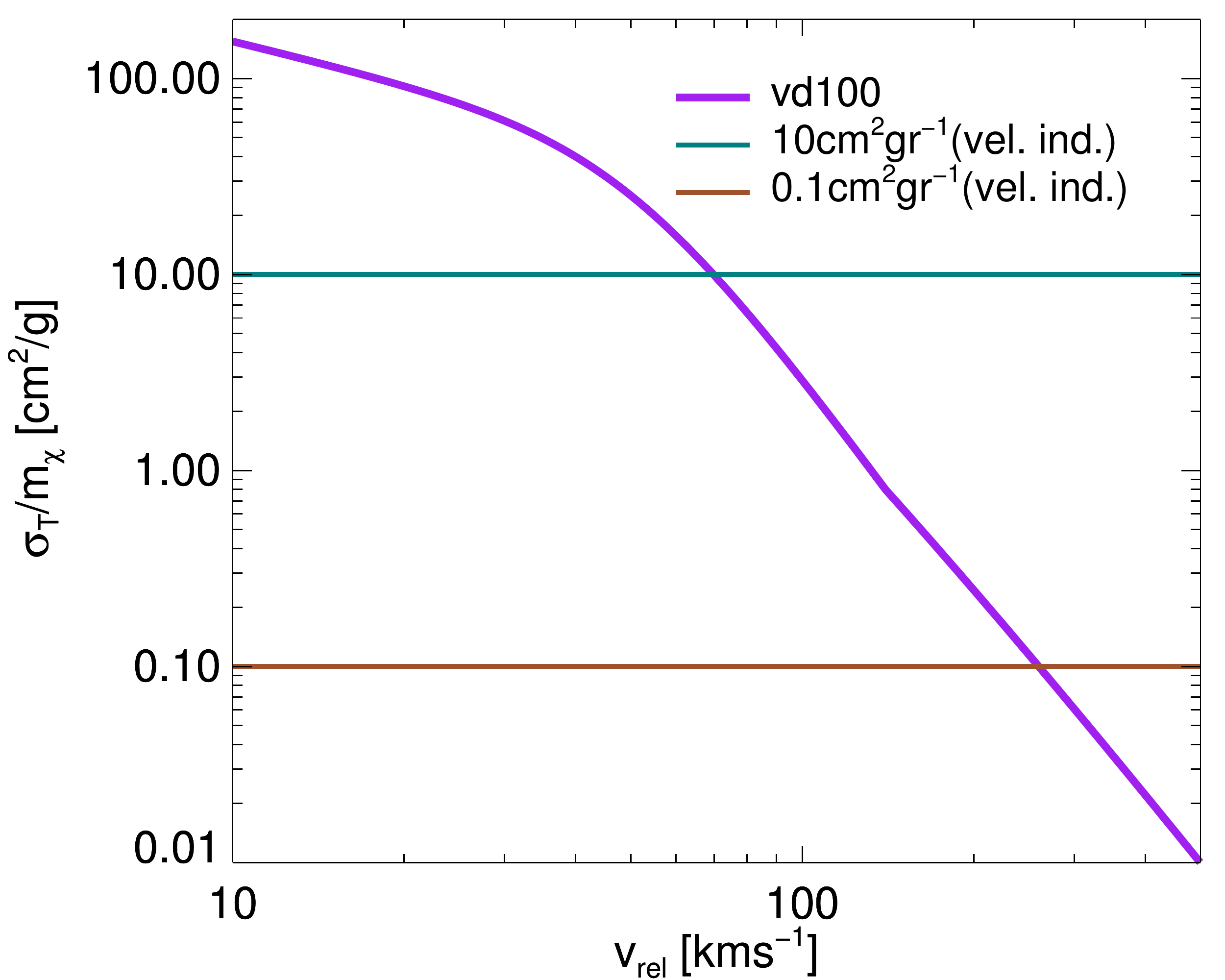}
    \caption{Presentation of the models used in this paper: CDM, 3.3~keV thermal relic-WDM, and vd100-SIDM (thick lines) compared to a series of alternative models (thin) lines. {\it Left-hand panel:} linear matter power spectra for CDM (black), thermal relic WDM models with masses 3.3, 2.5 and 7.3~keV (red, blue and green lines respectively), and the 7.1~keV mass, resonantly produced sterile neutrino models that bracket the region consistent with a dark matter decay interpretation of the reported 3.55~keV line: lepton asymmetry $L_6=9$ (orange dashed line) and $L_{6}=11$ (orange dotted line) with the region between highlighted in yellow. The 2.5 and 7.3~keV curves reflect the approximate 20:1 likelihood limits derived in \citet{Enzi21} and \citet{Nadler21} respectively. We include the half-mode mass, $M_\rmn{hm}$, for each model in the figure legend. {\it Right-hand panel:} the self-scattering transfer cross-section used in this paper, vd100 (purple), as a function of relative velocity, plus two velocity independent SIDM models: 10~cm$^2$g$^{-1}$ (teal) and 0.1~cm$^2$g$^{-1}$ (brown).}
    \label{fig:models}
\end{figure*}

The COCO-WDM simulation is identical to the CDM version except that it was performed assuming a thermal relic WDM particle of mass 3.3~keV, which corresponds to a half-mode mass scale, $M_\rmn{hm}=3.5\times10^{8}$~$\msun$ \citep{Bose16a}. This model was selected to be in mild tension with the Lyman-$\alpha$ forest constraint of \citet{Viel13}. Its linear matter power spectrum is a good approximation to that of the 7.1~keV-mass sterile neutrino that can explain the X-ray 3.55~keV line reported in M31 and galaxy clusters \citep{Boyarsky14a,Bulbul14} for a lepton asymmetry $L_6=10$ \citep{Lovell20}; note that the sterile neutrino is produced via the freeze-in mechanism, and therefore evades the nucleosynthesis constraints that limit the masses of true thermal relics to $>400$~keV \citep{Sabti20}. The 3.3~keV thermal relic power spectrum is in good agreement with the conservative constraints of \citet{Newton21} and \citet{Enzi21}, who determined a 20:1 likelihood ratio limit of 2.5~keV, but is in strong tension with the analysis of \citet{Nadler21} who instead reported a 20:1 likelihood ratio limit of 7.4~keV. Both limits are driven by MW satellite counts. We present the linear matter power spectra for these three thermal relic masses plus the power spectra for the 3.55~keV line-compliant sterile neutrino model in the left-hand panel of Fig.~\ref{fig:models}.    

For our SIDM results we consider the extreme velocity-dependent cross-section model and accompanying simulation from \citet{Zavala19a} and \citet{Turner21}. This model features elastic collisions with a Yukawa-like potential. The self-scattering transfer cross-section per unit mass, $\sigma_\rmn{T}/m_\rmn{X}$, is 100~cm$^2$g$^{-1}$ at $V_\rmn{rel}=14$~$\kms$, and 2~cm$^2$g$^{-1}$ at $V_\rmn{rel}=100$~$\kms$; we present this model transfer cross-section as a function of $V_\rmn{rel}$ in the right-hand panel of Fig~\ref{fig:models} alongside two velocity independent models: 10~cm$^2$g$^{-1}$ as the cross-section at which gravothermal collapse begins to occur, and 0.1~cm$^2$g$^{-1}$ for the cross-section where self-interactions are too rare to impact the halo structure. The simulation cosmological parameters are the same as in the original Aquarius runs \citep{Springel08b} and are thus consistent with the WMAP-1 cosmology \citep{wmap1}. The dark matter particle mass is $4.9\times 10^{4}$~$\msun$ and the Plummer equivalent gravitational softening length is 120.5~pc. Haloes and subhaloes were identified with the {\sc subfind} algorithm \citep{Springel01}. Throughout this paper, we refer to this SIDM model as vd100 and to the accompanying \citet{Zavala19a} simulation as AqA3-vd100.

\subsection{Host halo and subhalo selections}
\label{subsec:halosel}

Our goal is to select MW-analogue host haloes and then identify their subhaloes that could plausibly host satellite galaxies, taking into account the possibility that the stripping and destruction of satellites is not captured accurately by $N$-body simulations. We first present our host halo selection and then present the method for identifying satellites.

One of the uncertainties highlighted above with regard to whether a given model and host halo combination can fit the MW satellite mass distribution is the mass of the MW host halo itself. This mass is estimated to be in the range $M_\rmn{200,MW}\sim[1.0,1.4]\times10^{12}$~$\msun$ \citep{Callingham19}; $M_{200}$ is defined as the mass enclosed in a radius of mean density $200$ times the critical density for collapse, and is sometimes referred to as the virial mass. In order to investigate the effect of host mass we select a much broader mass range for potential hosts, which we discuss below. We start by choosing a host halo mass definition that will enable us to select MW-analogues in systems where the MW is located in the same FoF group as an M31 mass companion. For this purpose we use the gravitationally bound mass as determined by {\sc subfind}, which we denote as the dynamical mass, or $M_\rmn{D}$, at $z=0$. In addition to this dynamical mass, we will set a limit of the parent FoF virial mass, $M_\rmn{200,FoF}$, to be no more than the mass of the Local Group, and also set a distance limit to other massive haloes to avoid merging systems. Our numerical requirements are:

\begin{itemize}
    \item  $M_\rmn{D}=[0.5,1.8]\times10^{12}$~$\msun$.
    \item $M_\rmn{200,FoF}<3\times10^{12}$~$\msun$.
    \item Distance to any halo/subhalo with $M_\rmn{D}>2\times10^{11}$~$\msun$ is $>500$~kpc.
\end{itemize}

\noindent
We apply this procedure to both the CDM and WDM simulations, and identify 102 host systems that satisfy these criteria. Note that this method can select subhaloes that are not the largest structure in the parent FoF group. 

The assignment of a halo mass value to compare with observations is complicated by the presence of the MW within the Local Group, as hinted at above. The typical definition for the MW host mass is $M_{200}$, but if the MW is sufficiently close to the M31 galaxy that the two haloes effectively overlap, the true value of $M_{200}$ will include the contribution of both haloes. On the other hand, $M_\rmn{dyn}$ omits the contribution of substructure and can also vary depending on what material the halo finder considers to be gravitationally bound. We therefore develop a pseudo-$M_{200}$ for our haloes, which is defined as follows. We draw two radii around the centres-of-potential of each halo, 50 and 100~kpc, and compute the total mass within these two radii, including material that is both bound and unbound to the host smooth halo. We then calculate the NFW profile that joins the enclosed masses at 50 and 100~kpc, and finally compute the value of $M_{200}$ for this NFW. We label this mass $M_\rmn{200,H}$, and will use this as the mass definition for our haloes. 

For our satellite populations we select all subhaloes that could host MW satellites, irrespective of whether they are likely to be resolved in COCO with $>20$ particles at $z=0$. We employ merger trees to identify all subhaloes that have a peak dynamical mass of $M_\rmn{D,peak}>10^{8}$~$\msun$ and whose descendant at $z=0$ is within 300~kpc of one of the 102 host systems. The descendant may either be a surviving subhalo or the most bound particle associated with the halo before it was disrupted by the host halo.

In order to host a luminous galaxy, the halo must be massive enough to retain its gas and allow the cooling of that gas into stars. Prior to reionization cooling can happen in haloes massive enough to sustain atomic hydrogen cooling; the redshift at which the mass threshold for atomic cooling occurs is labelled $z_\rmn{cool}$. After reionization, the gas is heated to $2\times10^{4}$~K and the required halo virial temperature/mass for collapse increases, and occurs at what \citet{BenitezLlambay20} term the `critical mass'; we label the redshift at which the critical mass threshold is surpassed $z_\rmn{crit}$. In order to combine these two thresholds, we define a `condensation redshift', $z_\rmn{cond}$, which encodes the first redshift at which each halo can form stars:

    \begin{equation}
    z_\rmn{cond}=
\begin{cases} 
      z_\rmn{cool} & z_\rmn{cool}\ge 6, \\
      z_\rmn{crit} & z_\rmn{cool}<6. \\ 
   \end{cases}
   \label{eqn:zcond}
    \end{equation}

Subhaloes that never attain either mass threshold are considered to be completely dark and are thus removed from the sample. Finally, we identify the subhalo peak value of $M_{200}$ across time and use this as our primary definition of subhalo pre-infall mass; we refer to this pre-infall subhalo mass as $M_\rmn{200,i}$. 

Some fraction of satellites will have been destroyed by the present day as a result of either merging under dynamical friction or interactions with the strong tidal fields of the MW halo and disc \citep{GarrisonKimmel19}. In order to infer which subhaloes would have completely merged with their hosts prior to the present day, we adopt the merger time formula of \citet{Lacey16} at the time of infall, which we define as the first time at which the subhalo passes within the virial radius, $r_{200}$ of its $z=0$ host\footnote{We adopt a radius for the MW satellite systems of 300~kpc, which is larger than $r_{200}$ for an MW-analogue host. Therefore, by this definition some satellite haloes never undergo `infall'.}:

\begin{equation}
    t_\rmn{merge} = \frac{f(\epsilon)}{2C}\frac{M_\rmn{h}}{M_\rmn{sat}}\frac{1}{\ln{(1+M_\rmn{h}/M_\rmn{sat}})} \left(\frac{r_\rmn{circ}}{r_\rmn{200}}\right)^{1/2} \tau_\rmn{dyn},
    \label{eqn:tmerge}
\end{equation}

\noindent where $\tau_\rmn{dyn}$ is the dynamical time, defined as the ratio of the host virial radius to the host virial circular velocity, or $\tau_\rmn{dyn}=r_{200}/V_{200}$; the constant $C=0.43$, $\epsilon$ is the ratio of the measured orbital angular momentum to the angular momentum of a spherical orbit at that radius, and the function $f(\epsilon) = 0.90\epsilon^{0.47}+0.60$. $r_\rmn{circ}$ is the radius of the circular orbit that has the same energy as the measured subhalo orbit. 

In order to account for disruption by the disc, we consider the result of \citet{Grand21}, who found that the probability of satellite disruption flips from below 50~per~cent to above 50~per~cent around the 10~kpc initial pericentre mark; we thus eliminate from our sample all subhaloes for which the first pericentre impact parameter is $<10$~kpc\footnote{The stellar mass of this simulated MW-analogue is $6.7\times10^{10}$~$\msun$, and thus 20~per~cent more massive than the MW stellar mass as measured by \citet{Cautun20}, it may therefore be the case that our criterion is at the more aggressive end for destroying satellites.}. We estimate the distance of first pericentre from the host halo centre using a simplified version of the interpolation developed in \citet{Richings20} as follows. We identify the two simulation snapshots that bracket the time of pericentre; compute the two spherically averaged gravitational potentials and calculate their average value between the two snapshots; integrate the subhalo orbit in this potential both forwards from the first snapshot then backwards from the second; and finally compute the average pericentre over the two orbits. We differ from \citet{Richings20} only in that we use a spherically averaged potential rather than their anisotropic potential.  

Finally, we also generate an additional selection of $z=1$ independent haloes in order to check for the impact of using our infall mass functions. We adopt $z=1$ for our selection as a balance between haloes that can, on the one hand, form late and thus have a chance of obtaining a low, Crater~II-like density, and on the other hand can fall into the MW early enough to undergo significant stripping. In this selection we only include haloes with $M_{200}>10^{9}$~$\msun$. 

\subsection{Subhalo density profile fitting}

The resolution of the COCO simulations is only sufficient to resolve the half-light radii of the largest observed satellites ($\ge1$~kpc), and small, artificial density cores enhance the rate at which they are stripped. We therefore fit NFW profiles to each subhalo's measured density profile prior to infall, with a fitting radius range of $[r_\rmn{Pow},3r_\rmn{Pow}]$, where $r_\rmn{Pow}$ is the Power radius for convergence of the density profile \citep{Power03,Springel08b}. We first compute the circular velocity profile as a function of radius, $V_\rmn{c}(r)$. Our initial iteration of the fit then adopts the measured subhalo the maximum value of the $V_\rmn{c}$ curve, $V_\rmn{max}$, and the radius at which this maximum occurs, $r_\rmn{max}$, to infer values for the NFW fitting parameters, the characteristic radius $r_\rmn{s}$ and characteristic density $\rho_\rmn{s}$. We compute a quality of fit statistic $Q$ for a grid of parameters that span a factor of two either side of the first iteration values, where $Q$ takes the form:

\begin{equation}
    Q = 1/n\sum_{i}^{n}[\log{(\rho_\rmn{fit}(r_{i}))}-\log{(\rho_\rmn{sim}(r_{i}))}]^2,
\end{equation}

\noindent
 for density profile fit $\rho_\rmn{fit}(r_{i})$ and measured simulation density profile $\rho_\rmn{sim}(r_{i})$ over $n$ radius bins. We then select the parameters that minimize the value of $Q$ for each subhalo; the average quality of fit is the same for CDM and WDM. We reject haloes for which $Q>2$ as being possibly unrelaxed haloes or spurious results from the halo finder.  We then use these fits to infer the density profile and circular velocity profile to radii of $\ge$10~pc. 
 
 We have repeated this fitting procedure -- and subsequently the rest of the analysis in this paper -- with the Einasto profile \citep{Einasto65} of fixed shape parameter $\alpha=0.18$, which has been shown to provide a superior fit to the NFW profile for subhaloes \citep{Springel08b}. The results obtained using the Einasto profile do not show a systematic deviation from those obtained when using the NFW profile, therefore we only use NFW-derived results for the remainder of this paper.

In order to generate predictions for the vd100 SIDM model, we adapt the COCO-CDM subhalo sample with input from the AqA3-vd100 simulation; we therefore assume that the primordial density field and subsequent halo assembly history are not affected by self-interactions. We adopt the COCO-CDM NFW-fits as a starting point and then impose cores in the following manner. We define a core radius, $r_\rmn{core}$, as the radius at which the density of a core is half the density of the fitted NFW profile. In the AqA3-vd100 simulation, we compute this core radius for isolated haloes and find that, for haloes in which a core can be measured, $r_\rmn{core}$ is approximately 23~per~cent of the measured halo $r_\rmn{max}$ for $M_{200}=10^{8}$~$\msun$ haloes and 18~per~cent of the measured $r_\rmn{max}$ for $M_{200}=10^{10}$~$\msun$ haloes. We adopt $r_\rmn{core}=0.2\times r_\rmn{max}$ for all subhaloes prior to gravothermal collapse, which is slightly larger than for the majority of haloes, in order to compensate for changes to the $V_\rmn{max}$--$r_\rmn{max}$ relation from CDM to SIDM which we describe below.

The AqA3-vd100 also predicts a marginally different $M_{200}$--$r_\rmn{max}$ relation to CDM, with at $r_\rmn{max}\sim30$~per~cent larger at fixed $M_{200}$ for AqA3-vd100 haloes than for the CDM haloes in the original AqA3 simulation; $V_\mathrm{max}$ is instead suppressed at the percent level. We will take this account into effect as part of the following computation of non-gravothermally collapsed SIDM profiles. First, we retain the COCO-CDM NFW profile at radii $r>\beta r_\rmn{max}$, where $r_\rmn{max}$ is still the CDM-NFW-derived value and $\beta$ is a fitting parameter that we use to obtain the approximate required $V_\mathrm{max}$--$r_\mathrm{max}$ relation. We enforce a flat density profile for $r<r_\rmn{core}$ with an amplitude of $0.5\rho_\rmn{fit}(r_\rmn{core})$, and then bridge the $r_\rmn{core}<r<\beta r_\rmn{max}$ range with a function of the form:

\begin{equation}
    \rho = \frac{\rho_\rmn{i}}{(1+r/r_\rmn{i})^{\gamma}},
    \label{eqn:bridgefit}
\end{equation}

\noindent where the index $\gamma$, scale radius $r_\rmn{i}$ and characteristic density $\rho_\rmn{i}$ are set to ensure that $\rho(\beta r_\rmn{max})=\rho_\rmn{NFW}(\beta r_\rmn{max})$, the density at $r_\rmn{core}$ in equation~\ref{eqn:bridgefit} is equal to the core density derived above, and the mass in the equation~\ref{eqn:bridgefit} is maximized. Finally, we determine that the AqA3-vd100 $V_\rmn{max}-r_\mathrm{max}$ is met for $\beta=1.7$, therefore we adopt this value of $\beta$ for all our SIDM haloes. With the accompanying change to $r_\rmn{max}$, $r_\rmn{core}\sim0.15r_\rmn{max}$. Finally, we assume that baryonic physics do not have any effect on the density profile shape. This is likely a good approximation for faint dSphs where the mass-to-light ratio is $>1000$ \citep{Read19}, but not for the Magellanic Clouds: we therefore take care not to overinterpret our findings for these two objects.

\subsection{SIDM gravothermal collapse}

It has been argued that the SIDM model fails to generate subhaloes of sufficient density to host the high-velocity dispersion ultrafaint dwarf galaxies \citep{Zavala19a,Silverman23}. In response, \citet{Zavala19a} showed that in extreme SIDM models haloes undergo gravothermal collapse to a degree that matches the densest satellites; we therefore need to account for this process in our analysis. 

\begin{figure*}
    \centering
    \includegraphics[scale=0.35]{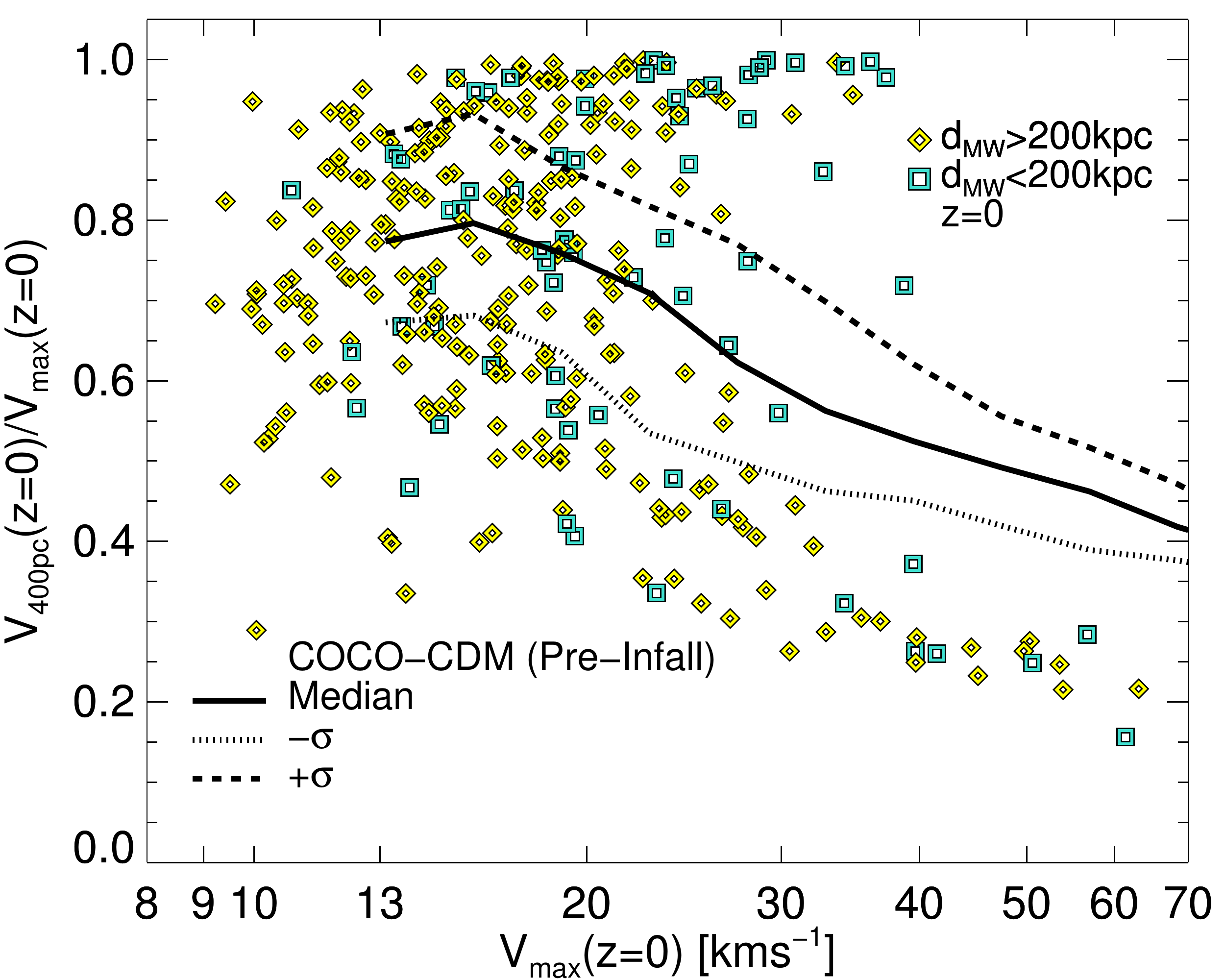} \includegraphics[scale=0.35]{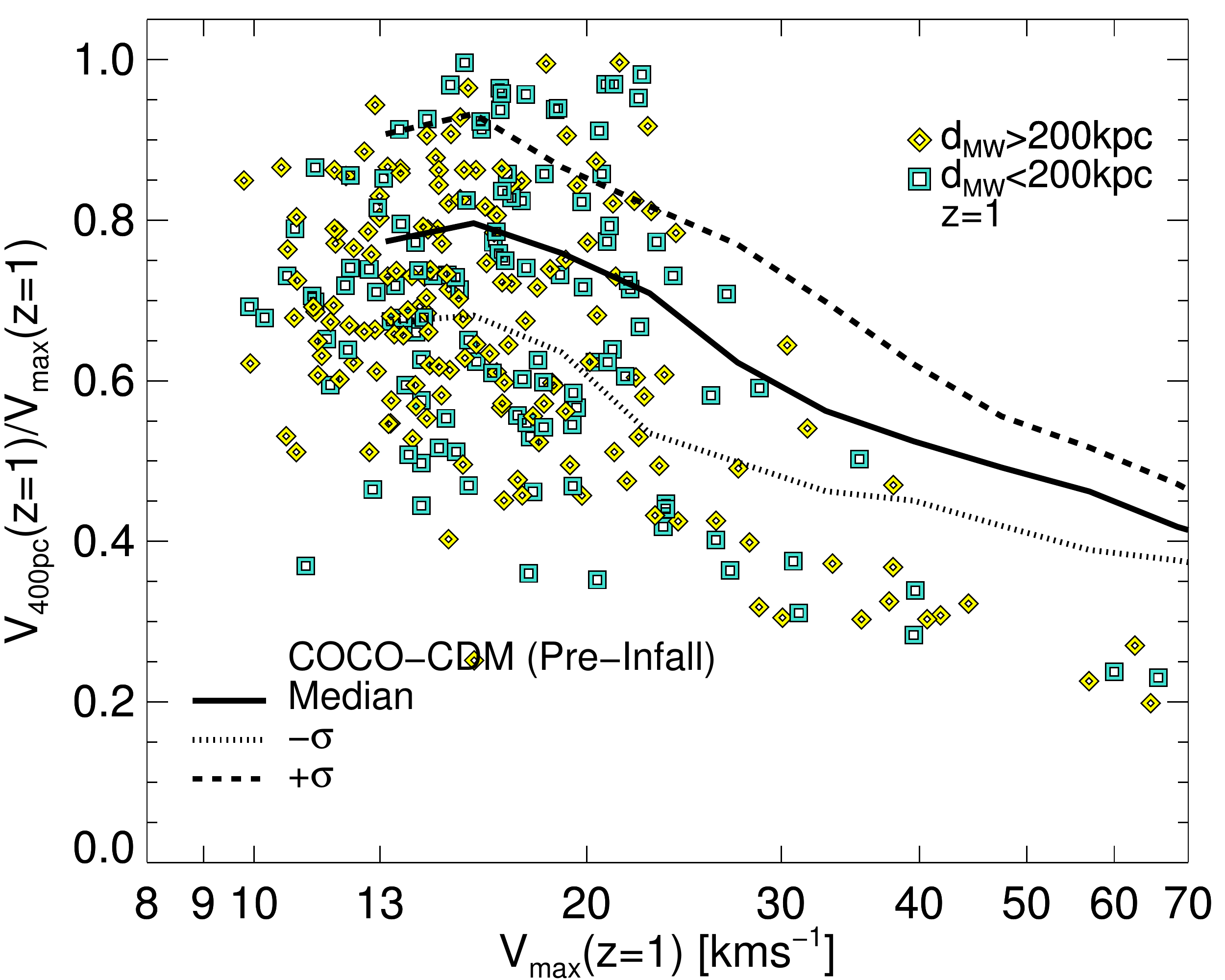}
    \caption{The ratio of halo/subhalo circular velocity measured at 400~pc to $V_\rmn{max}$ as a function of $V_\rmn{max}$ for AqA3-vd100 haloes \citep{Zavala19a,Turner21} at $z=0$ (left-hand panel) and $z=1$ (right-hand panel). Objects within 200~kpc (physical) of the MW-analogue halo centre are shown as green squares and objects outside this radius are shown as yellow diamonds. The solid black line indicates the ratio between the quantities for the COCO-CDM profile with the median $V_\rmn{max}$--$V_\rmn{400pc}$ relation measured from the COCO-CDM satellite selection, and the dotted and dashed lines delineate the 68~per~cent regions for the COCO-CDM data.}
    \label{fig:GCz0z1}
\end{figure*}

In order to infer possible density profiles for gravothermally collapsed haloes in SIDM, we first study the population of haloes and subhaloes in the AqA3-vd100 simulation and then draw expectations for gravothermally collapsed halo distributions in our COCO-SIDM model. We characterize each profile using the value of the circular profile at two radii: $r_\rmn{max}$ plus and 400~pc, with the circular velocity values labelled $V_\rmn{max}$ and $V_\rmn{400pc}$ respectively. We choose 400~pc as this is the smallest radius that can plausibly be resolved in the AqA3-vd100 simulation. We plot the ratio $V_\rmn{400pc}/V_\rmn{max}$ as a function of $V_\rmn{max}$ in Fig.~\ref{fig:GCz0z1}, highlighting the value of this ratio relative to the $V_\rmn{400pc}/V_\rmn{max}$ ratio for our COCO-CDM MW satellite halo selection\footnote{We employ the satellite halo selection rather than the $z=1$ isolated halo selection -- or a notional $z=0$ isolated halo selection -- because it contains haloes in the $[10^{8},10^{9}]$~$\msun$ mass range. We have checked that this median relation is consistent with the $V_\rmn{400pc}/V_\rmn{max}$ predictions inferred from the \citet{Neto07} mass--concentration relation.} and including the distance of each halo to the centre of the MW-analogue halo. 

At $z=0$ and $z=1$ there is a wide variety of velocity ratio values as a function of $V_\rmn{max}$. The cored haloes are defined as having sub-CDM densities. By this definition, cores typically occur for haloes with $V_\rmn{max}>30$~$\kms$, and towards smaller velocities the proportion of haloes with CDM-value ratios or even super-CDM ratios increases, the latter defined as objects with $V_\rmn{400pc}$ greater than 84~per~cent of the CDM haloes at a given $V_\rmn{max}$ (or $M_{200}$, see below). The proportion of super-CDM density satellites with $V_\mathrm{max}>14$~$\kms$ increases from 9~per~cent at $z=1$ to 28~per~cent by $z=0$, thus indicating that 66~per~cent of the collapse after $z=1$. At $z=1$ there is no striking evidence for any difference in the profile distributions between haloes that are within $200$~kpc of the central MW-analogue halo to those outside this range. There is weak evidence that haloes closer to the MW-analogue at $z=0$ may have undergone further collapse as anticipated in \citet{Nishikawa20}, particularly for super-CDM objects with $V_\rmn{max}>30$~$\kms$; this is a population of objects that is also absent at $z=1$. Finally, we emphasize here that care must be taken when using $V_\rmn{max}$ as a proxy for subhalo mass in these simulations. At fixed $V_\rmn{max}$, the mass within $r_\rmn{max}$ is proportional to $r_\rmn{max}$, so if cored objects with $30$~$\kms$ have $r_\rmn{max}\sim4$~kpc, and the size of the collapsed haloes of the same $V_\rmn{max}$ is 400~pc, the masses differ by a factor of 10.

We have shown that the gravothermal collapse occurs in large part between $z=1$ and $z=0$, and correlates weakly with distance to the host galaxy. Our goal is to turn this information into a scheme by which to infer which SIDM subhaloes gravothermally collapse and subsequently what the profile might be. From Fig.~\ref{fig:GCz0z1} we can infer that the collapse will sometimes occur at the same time that the subhalo is stripped by the host, independent of expectation that infall could hasten any collapse \citep{Nishikawa20}. A definitive analysis would treat these two processes together, and is beyond this work. Instead, we identify isolated haloes at $z=1$ and analyse their inner profiles at $z=0$, assuming that the role of stripping is negligible. This assumption is supported by the recent result of \citet{ZengZ22} who found that stripping does not affect the collapse; we will perform a first-order test of this hypothesis in Section~\ref{sec:indsats}.

\begin{figure}
    \centering
    \includegraphics[scale=0.345]{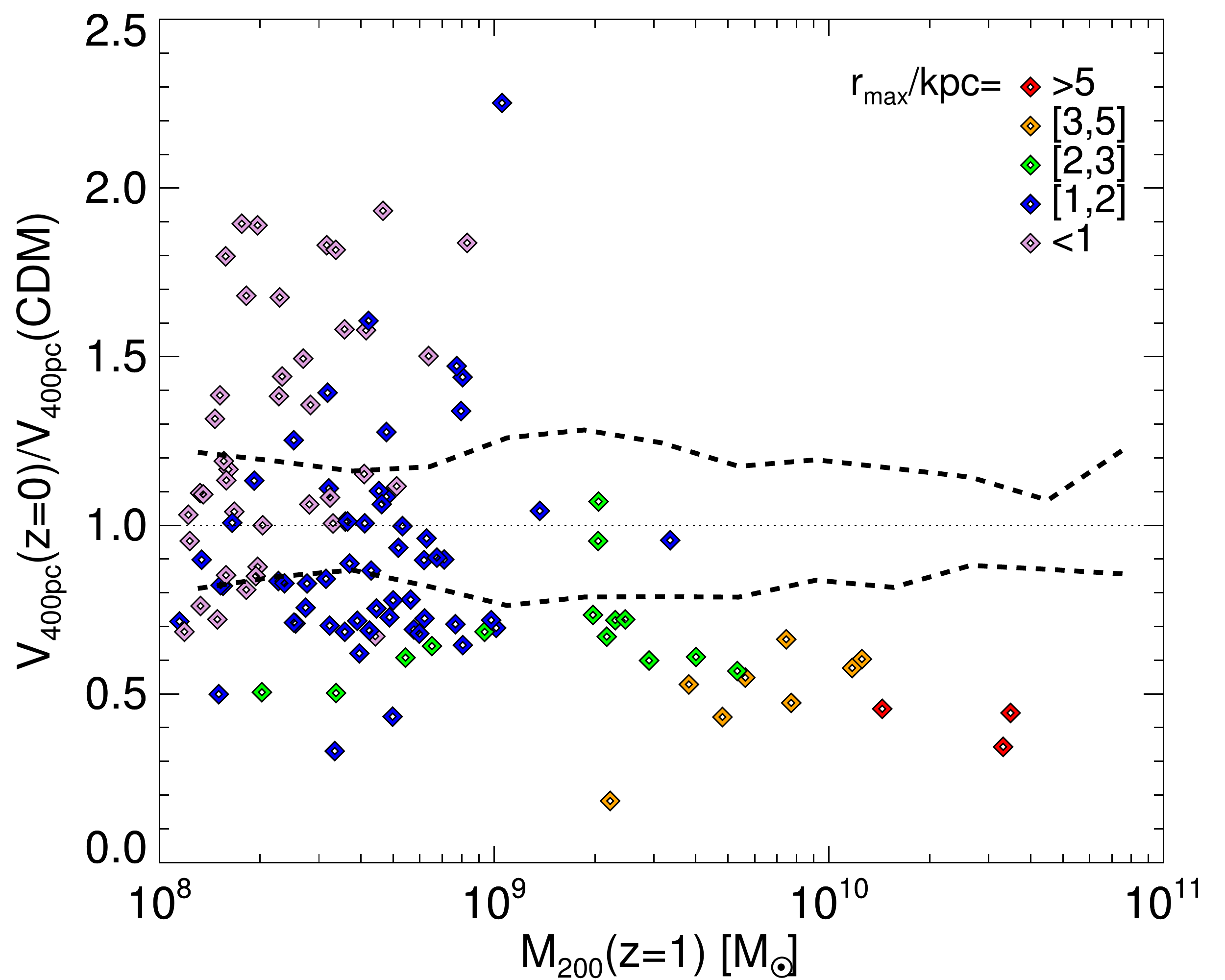} 
    \caption{The ratio of the $z=0$ AqA3-vd100 halo circular velocity measured at 400~pc ($V_\mathrm{400pc}$) to the median CDM-COCO $V_\mathrm{400pc}$ measured at the same radius, as a function of the $z=1$ $M_{200}$ for haloes that are isolated at $z=1$. The symbol colour indicates the $r_\rmn{max}$ measured at $z=1$. We include all isolated $z=1$ haloes, irrespective of their density profile shape at $z=1$. The pair of dashed lines indicate the 68~per~cent scatter in the COCO-CDM relation.}
    \label{fig:v4vCrat}
\end{figure}

In order to infer the change in circular velocity profile due to collapse, we measure the difference between the measured 
$z=0$ circular velocity at 400~pc and the COCO-CDM value at the same radius, as a function of each halo's $M_{200}$ at $z=1$\footnote{We adopt $z=1$ as this was the earliest redshift snapshot that was written to disk.}. We select all of the AqA3-vd100-isolated haloes at $z=1$, compute the $V_\rmn{400pc}$ of their $z=0$ descendants and calculate the ratio of these measured $V_\rmn{400pc}$ to the $V_\rmn{400pc}$ inferred for a halo of this mass from the COCO-CDM relation described above, now applied for  $M_{200}$ instead of $V_\mathrm{max}$. We present the results in Fig.~\ref{fig:v4vCrat}. 

The collapse likelihood is strongly mass dependent, reflecting the strong velocity-dependence of the cross-section of the vd100 model \citep[see][]{Zavala19a}. From the definition of collapse as having $V_\rmn{400pc}$ values denser than the CDM 68~per~cent range, collapse only occurs for $M_{200}<10^{9}$~$\msun$, and this for only a minority of subhaloes in this mass range. For a less strict definition of collapse -- namely that the $V_\rmn{400pc}$ value is greater than the CDM mean -- we find that 50~per~cent of haloes with $M_{200}=3\times10^{8}$~$\msun$ have collapsed, with the collapse fraction increasing further towards lower masses. The distribution of circular velocity ratios is broad rather than bimodal, and may reflect a phase of the collapse that is gradual prior to the eventual instantaneous gravothermal catastrophe \citep{YangS22}. There are four objects with CDM-style densities at $10^{9}<M_{200}<4\times10^{9}$~$\msun$, otherwise all of these objects are cored. There is a trend for objects that have $r_\rmn{max}<1.5$~$\kms$ to be more likely to have collapsed; however, given the small number statistics, we do not use this finding in the remainder of our analysis. 

We now introduce our application of this result to infer gravothermally collapsed profiles for our haloes. In summary, we use the AqA3-vd100 data to infer new profiles for the COCO-SIDM haloes at the population level, assigning some proportion of haloes to retain their cored profiles, some proportion to have CDM profiles, and the remainder to have super-CDM density profiles.  

First, we restrict the possibility of collapse to haloes with $M_{200}<10^{9}$~$\msun$. For the AqA3-vd100 haloes in the Fig.~\ref{fig:v4vCrat} sample, we compute the cumulative distribution of $V_\rmn{400pc}$ ratios and then invert this distribution to infer a probability distribution for a given halo to obtain a given $V_\rmn{400pc}$. Turning to the COCO-SIDM haloes with $M_{200}<10^{9}$~$\msun$, we generate random values for the  $V_\rmn{400pc}$ ratio as drawn from the AqA3-vd100-derived distribution. If the value drawn is $<0.8$ -- i.e. less dense than the CDM range -- we retain the cored value, and if the value is in the range [0.8,1.2] -- in the CDM range -- we adopt the original COCO-CDM-NFW profile for the SIDM halo. If the selected ratio is $>1.2$ we infer a halo $V_\rmn{400pc}$ as the product of the randomly draw ratio and the $V_\rmn{400pc}$ of the COCO-CDM-NFW halo. We then draw a new NFW profile that has the same $M_\rmn{200}$ as the SIDM halo but also passes through the new $V_\rmn{400pc}$ value. Note that we neglect much of the evolution of collapsed fraction with mass: we will comment later in the text where this assumption will affect our results.  
 
 In conclusion, we have shown that in the vd100 model it is primarily $<10^{9}$~$\msun$ haloes that collapse and that most of this collapse occurs after $z=1$. We then derived an {\em ad hoc} model to infer a variety of density profiles for COCO-SIDM subhaloes. We argued that to first-order the impact of tides was weak and did not attempt to compute the collapse time for these haloes; we defer this analysis to future semi-analytical work \citep{Meshveliani22}. We note that our analysis is unable to capture the collapse likelihood of subhaloes that are accreted prior to $z=1$.
 
 \subsection{Observed satellite selection}
 
 The core goal of this project is to match subhaloes to observed MW satellite galaxies. We consider 39 MW satellites for which we obtained sizes and masses from the literature, defined as objects within 300~kpc of the MW centre that are considered to be DM dominated. We adopt deprojected 3D half-stellar mass radii and enclosed masses where available, and in cases where only projected 2D masses and sizes are published we obtain approximate deprojected radii and enclosed masses by multiplying the 2D radii by $4/3$ and the reported masses by $(4/3)^{3}$. Velocity dispersions are converted into circular velocities using the estimator of \citet{Wolf10}. We present the satellite stellar masses, deprojected half-light radii and circular velocities, present-day MW-distances and pericentric distances in Table~\ref{tab:obs}, together with the reference papers/compilations from which these properties were obtained/derived. We include uncertainties on $V_\rmn{h}$ and $d_\rmn{peri}$ but not on the other three quantities. A complete analysis would include the uncertainty on the half-light radii on finding matches to satellites; for simplicity we neglect this consideration here, especially in the context of the many uncertainties involved elsewhere in this process.   

 \begin{table*}
     \centering
      \caption{Estimates for properties of satellites used in this paper: stellar mass $M_{*}$, 3D deprojected half-light radius $r_\rmn{h}$, 3D deprojected circular velocity $V_\rmn{h}$, present-day distance to the MW centre, $d_\rmn{MW}$, and pericentre distance to MW centre, $d_\rmn{peri}$. Errors are presented for $V_\rmn{h}$ and $d_\rmn{peri}$.  We do not include pericentre estimates for the Magellanic clouds. References:  $^{a}$ \citet{Torrealba19}; $^{b}$ \citet{Pace22}; $^{c}$ \citet{Torrealba16b}; $^{d}$ \citet{Wolf10}; $^{e}$ \citet{McConnachie12}; $^{f}$ \citet{Koch09}; $^{g}$ \citet{LiT18}; $^{h}$ \citet{Torrealba16}; $^{i}$ \citet{Caldwell17}; $^{j}$ \citet{Zoutendijk21}; $^{k}$ \citet{Koposov18}; $^{l}$ \citet{Buckley15}; $^{m}$ \citet{Richstein22}; $^{n}$ \citet{Cerny23}; $^{o}$ \citet{Fritz19}; $^{p}$ \citet{MutluPakdil18}; $^{q}$ \citet{Minor19}; $^{r}$ \citet{Caputo16}; $^{s}$ \citet{Gaia18}; $^{t}$ \citet{Walker16}; $^{u}$ \citet{Simon20}. The Sagittarius pericentre adopted from \citet{Gaia18} was derived from their MW potential model number 2.}

     \begin{tabular}{l|r|r|r|r|r}
Name & $M_{*}$~($\msun$) & $ r_\rmn{h}$~(kpc) & $V_\rmn{h}$~(kms$^{-1}$) & $d_\rmn{MW}$~(kpc) & $d_\rmn{peri}$~(kpc) \\
\hline
Antlia II  & $ 8.8\times10^{5} $$^{a}$ & $ 5.200 $$^{a}$ & $ 10.6^{+2.1}_{-2.1} $$^{a}$ & $ 130.0 $$^{a}$ & $ 38.2^{+10.0}_{-7.8}$$^{b}$ \\
Aquarius II  & $ 2.0\times10^{3} $$^{c}$ & $ 0.160 $$^{c}$ & $ 9.4^{+1.6}_{-1.6} $$^{c}$ & $ 107.9 $$^{c}$ & $ 55.1^{+40.8}_{-23.7}$$^{b}$ \\
Bootes I  & $ 2.9\times10^{4} $$^{d}$ & $ 0.322 $$^{d}$ & $ 17.8^{+3.8}_{-3.8} $$^{d}$ & $ 64.0 $$^{d}$ & $ 37.9^{+7.5}_{-6.8}$$^{b}$ \\
Bootes II  & $ 1.0\times10^{3} $$^{e}$ & $ 0.068 $$^{e}$ & $ 28.0^{+19.7}_{-19.7} $$^{f}$ & $ 40.0 $$^{e}$ & $ 35.8^{+2.9}_{-2.5}$$^{b}$ \\
Canes Venatici I  & $ 2.3\times10^{5} $$^{d}$ & $ 0.750 $$^{d}$ & $ 12.4^{+1.3}_{-1.3} $$^{d}$ & $ 218.0 $$^{d}$ & $ 84.5^{+53.6}_{-37.2}$$^{b}$ \\
Canes Venatici II  & $ 7.9\times10^{3} $$^{d}$ & $ 0.097 $$^{d}$ & $ 8.0^{+1.6}_{-1.6} $$^{d}$ & $ 161.0 $$^{d}$ & $ 47.5^{+46.8}_{-29.7}$$^{b}$ \\
Carina  & $ 3.8\times10^{5} $$^{d}$ & $ 0.334 $$^{d}$ & $ 11.1^{+0.5}_{-0.5} $$^{d}$ & $ 107.0 $$^{d}$ & $ 77.9^{+24.1}_{-17.9}$$^{b}$ \\
Carina II  & $ 5.4\times10^{3} $$^{g}$ & $ 0.091 $$^{g}$ & $ 6.9^{+0.3}_{-0.3} $$^{g}$ & $ 37.4 $$^{g}$ & $ 29.2^{+0.6}_{-0.6}$$^{b}$ \\
Carina III  & $ 7.8\times10^{2} $$^{g}$ & $ 0.030 $$^{g}$ & $ 9.7^{+3.6}_{-3.6} $$^{g}$ & $ 27.4 $$^{g}$ & $ 28.8^{+0.6}_{-0.6}$$^{b}$ \\
Coma Berenices  & $ 3.7\times10^{3} $$^{e}$ & $ 0.103 $$^{e}$ & $ 9.7^{+1.5}_{-1.5} $$^{e}$ & $ 45.0 $$^{e}$ & $ 42.5^{+1.6}_{-1.6}$$^{b}$ \\
Crater II  & $ 4.1\times10^{5} $$^{h}$ & $ 1.421 $$^{h}$ & $ 5.7^{+0.7}_{-0.7} $$^{i}$ & $ 117.5 $$^{h}$ & $ 24.0^{+5.6}_{-5.2}$$^{b}$ \\
Draco  & $ 2.9\times10^{5} $$^{d}$ & $ 0.291 $$^{d}$ & $ 17.7^{+1.3}_{-1.3} $$^{d}$ & $ 76.0 $$^{d}$ & $ 58.0^{+11.4}_{-9.5}$$^{b}$ \\
Fornax  & $ 2.0\times10^{7} $$^{d}$ & $ 0.944 $$^{d}$ & $ 18.4^{+0.5}_{-0.5} $$^{d}$ & $ 149.0 $$^{d}$ & $ 76.7^{+43.1}_{-27.9}$$^{b}$ \\
Grus I  & $ 6.3\times10^{2} $$^{j}$ & $ 0.064 $$^{j}$ & $ 18.0^{+8.8}_{-8.8} $$^{j}$ & $ 120.0 $$^{j}$ & $ 48.9^{+27.0}_{-22.9}$$^{b}$ \\
Hercules  & $ 3.7\times10^{4} $$^{d}$ & $ 0.300 $$^{d}$ & $ 10.4^{+2.2}_{-2.2} $$^{d}$ & $ 126.0 $$^{d}$ & $ 67.4^{+15.5}_{-16.1}$$^{b}$ \\
Hydra II  & $ 2.0\times10^{4} $$^{j}$ & $ 0.068 $$^{j}$ & $ 6.0^{+1.9}_{-1.9} $$^{j}$ & $ 151.0 $$^{j}$ & $ 99.2^{+30.6}_{-55.7}$$^{b}$ \\
Hydrus I  & $ 6.0\times10^{3} $$^{k}$ & $ 0.053 $$^{k}$ & $ 4.7^{+0.9}_{-0.9} $$^{k}$ & $ 28.0 $$^{k}$ & $ 45.8^{+16.1}_{-6.0}$$^{b}$ \\
LMC  & $ 1.5\times10^{9} $$^{l}$ & $ 8.000 $$^{l}$ & $ 80.0^{+10.0}_{-10.0} $$^{l}$ & $ 50.0 $$^{l}$ & $  --  $ \\
Leo I  & $ 5.5\times10^{6} $$^{d}$ & $ 0.388 $$^{d}$ & $ 15.7^{+0.8}_{-0.8} $$^{d}$ & $ 258.0 $$^{d}$ & $ 47.5^{+40.9}_{-24.0}$$^{b}$ \\
Leo II  & $ 7.4\times10^{6} $$^{d}$ & $ 0.233 $$^{d}$ & $ 11.6^{+0.0}_{-0.0} $$^{d}$ & $ 236.0 $$^{d}$ & $ 61.4^{+62.3}_{-34.7}$$^{b}$ \\
Leo IV  & $ 1.9\times10^{4} $$^{e}$ & $ 0.275 $$^{e}$ & $ 6.9^{+2.7}_{-2.7} $$^{e}$ & $ 155.0 $$^{e}$ & $ 66.8^{+60.7}_{-44.1}$$^{b}$ \\
Leo V  & $ 1.1\times10^{4} $$^{e}$ & $ 0.180 $$^{e}$ & $ 7.9^{+2.5}_{-2.5} $$^{e}$ & $ 179.0 $$^{e}$ & $ 165.8^{+5.8}_{-49.2}$$^{b}$ \\
Pegasus III  & $ 2.0\times10^{3} $$^{m}$ & $ 0.157 $$^{m}$ & $ 14.4^{+4.7}_{-4.7} $$^{m}$ & $ 215.0 $$^{m}$ & $ 141.0^{+87.8}_{-79.3}$$^{b}$ \\
Pegasus IV  & $ 4.4\times10^{3} $$^{n}$ & $ 0.041 $$^{n}$ & $ 5.7^{+1.9}_{-1.9} $$^{n}$ & $ 90.0 $$^{n}$ & $ 32.0^{+18.0}_{-14.0}$$^{n}$ \\
Phoenix II  & $ 1.4\times10^{3} $$^{o}$ & $ 0.049 $$^{o}$ & $ 25.3^{+14.4}_{-14.4} $$^{p}$ & $ 80.0 $$^{o}$ & $ 84.6^{+91.3}_{-35.6}$$^{b}$ \\
Pisces II  & $ 8.6\times10^{3} $$^{g}$ & $ 0.077 $$^{g}$ & $ 10.3^{+3.5}_{-3.5} $$^{m}$ & $ 181.0 $$^{g}$ & $ 130.5^{+70.1}_{-73.3}$$^{b}$ \\
Reticulum II  & $ 2.4\times10^{3} $$^{q}$ & $ 0.055 $$^{q}$ & $ 5.6^{+0.1}_{-0.1} $$^{q}$ & $ 30.0 $$^{q}$ & $ 37.0^{+2.9}_{-5.3}$$^{b}$ \\
SMC  & $ 4.8\times10^{8} $$^{r}$ & $ 3.000 $$^{r}$ & $ 58.0^{+5.0}_{-5.0} $$^{r}$ & $ 61.0 $$^{r}$ & $  --  $ \\
Sagittarius  & $ 2.1\times10^{7} $$^{e}$ & $ 3.449 $$^{e}$ & $ 23.7^{+1.1}_{-1.1} $$^{e}$ & $ 18.0 $$^{e}$ & $ 14.3^{+2.3}_{-2.3}$$^{s}$ \\
Sculptor  & $ 2.3\times10^{6} $$^{d}$ & $ 0.375 $$^{d}$ & $ 16.1^{+0.5}_{-0.5} $$^{d}$ & $ 86.0 $$^{d}$ & $ 44.9^{+4.3}_{-3.9}$$^{b}$ \\
Segue I  & $ 3.4\times10^{2} $$^{e}$ & $ 0.039 $$^{e}$ & $ 8.3^{+1.6}_{-1.6} $$^{e}$ & $ 28.0 $$^{e}$ & $ 19.8^{+4.2}_{-4.8}$$^{b}$ \\
Segue II  & $ 8.6\times10^{2} $$^{e}$ & $ 0.046 $$^{e}$ & $ 7.1^{+1.6}_{-1.6} $$^{e}$ & $ 41.0 $$^{e}$ & $ 18.0^{+3.8}_{-3.1}$$^{b}$ \\
Sextans  & $ 4.4\times10^{5} $$^{d}$ & $ 1.019 $$^{d}$ & $ 12.1^{+0.9}_{-0.9} $$^{d}$ & $ 89.0 $$^{d}$ & $ 82.8^{+3.2}_{-4.9}$$^{b}$ \\
Tucana II  & $ 1.4\times10^{3} $$^{t}$ & $ 0.220 $$^{t}$ & $ 11.2^{+2.7}_{-2.7} $$^{t}$ & $ 57.0 $$^{t}$ & $ 44.8^{+12.3}_{-10.1}$$^{b}$ \\
Tucana IV  & $ 7.0\times10^{2} $$^{u}$ & $ 0.169 $$^{u}$ & $ 11.5^{+2.9}_{-2.9} $$^{u}$ & $ 47.0 $$^{u}$ & $ 32.1^{+18.5}_{-12.6}$$^{b}$ \\
Ursa Major I  & $ 1.4\times10^{4} $$^{e}$ & $ 0.425 $$^{e}$ & $ 16.2^{+2.2}_{-2.2} $$^{e}$ & $ 102.0 $$^{e}$ & $ 49.9^{+47.2}_{-15.6}$$^{b}$ \\
Ursa Major II  & $ 4.1\times10^{3} $$^{e}$ & $ 0.199 $$^{e}$ & $ 14.1^{+2.6}_{-2.6} $$^{e}$ & $ 38.0 $$^{e}$ & $ 41.4^{+3.4}_{-3.6}$$^{b}$ \\
Ursa Minor  & $ 2.9\times10^{5} $$^{d}$ & $ 0.588 $$^{d}$ & $ 20.2^{+1.3}_{-1.3} $$^{d}$ & $ 78.0 $$^{d}$ & $ 55.7^{+8.4}_{-7.0}$$^{b}$ \\
Willman I  & $ 1.0\times10^{3} $$^{e}$ & $ 0.033 $$^{e}$ & $ 9.1^{+2.4}_{-2.4} $$^{e}$ & $ 43.0 $$^{e}$ & $ 16.2^{+5.2}_{-3.0}$$^{b}$ \\

     \end{tabular}
    
     \label{tab:obs}
 \end{table*}
 
\subsection{Stripping procedure}

The principle behind our stripping procedure is to apply the \citet{Errani21} algorithm for generating stripped profiles from an input, unstripped NFW profile. We perform this procedure for each combination of fitted subhalo and observed satellites to determine a stripped density profile, $\rho_\rmn{st}$, and adopt a circular velocity profile, $V_\rmn{st}$ that is consistent with the measured satellite's deprojected half-light radius, $r_\rmn{h}$, and the $1\sigma$ uncertainty on the circular velocity at $r_{200}$, $V_\rmn{h}$. For haloes in which the unstripped value of the circular velocity at radius $r_\rmn{h}$ is below 5~per~cent of the lower bound on the observed $V_\rmn{h}$, it is not possible to fit the satellite with that particular subhalo. In all other cases, we apply the \citet{Errani21} fitting formula:

\begin{equation}
    \rho_\rmn{st} = \rho_\rmn{fit}\times \frac{\exp{(-r/r_\rmn{cut})}}{(1+r_\rmn{s}/r_\rmn{cut})^{\kappa}},
\end{equation}

\noindent where $\rho_\rmn{fit}$ is the pre-infall fitted density profile, $r_\rmn{s}$ is the scale radius of the initial NFW profile, and $\kappa$ is a fitting parameter that takes the value $\kappa=0.3$. $r_\rmn{cut}$ is the truncation radius, which is given by:

\begin{equation}
    r_\rmn{cut} = 0.44r_\rmn{max,0}\times\left(\frac{M_\rmn{max}}{M_\rmn{max,0}}\right)^{0.44}\left[1-\left(\frac{M_\rmn{max}}{M_\rmn{max,0}}\right)^{0.3}\right]^{-1.1},
\end{equation}

\noindent with $M_\rmn{max,0}$ the mass within $r_\rmn{max,0}$ of the unstripped profile and $M_\rmn{max}$ the mass with the $r_\rmn{max}$ of the stripped profile; $M_\rmn{max}$ is the parameter we fit to obtain the profile that matches the observed satellite masses and radii. Given that the $1\sigma$ uncertainty in $V_\rmn{h}$ allows for a range of $V_\rmn{st}$ profiles, where possible we select the profile with the highest value of $V_\rmn{h}$ that satisfies the tidal radius criterion discussed in the following subsection. This process was developed using NFW haloes and not cored or gravothermally collapsed profiles. It is likely that the rate of stripping is stronger for cored profiles than for cusps (Errani et al. in preparation); nevertheless, we proceed to treat the algorithm as equally accurate for our cored SIDM profiles as for cusps in this paper, and defer a detailed comparison to future work.   

\subsection{Viability of matches given orbital properties}

We have thus far developed an algorithm to ascertain the degree of stripping, if any, that must be applied to a given subhalo in order to fit the internal kinematics of an observed satellite galaxy. This procedure assumes that this amount of stripping can be achieved without reference to the subhalo orbit, which is a crucial factor in determining whether the required degree of stripping is viable. For example, if a given subhalo--satellite combination requires that the subhalo lose 90~per~cent of its mass within $r_\rmn{h}$, the subhalo will have had to have passed well within $100$~kpc of the host halo centre to effect this degree of stripping. A comprehensive set of matching criteria must therefore ask not just whether a satellite can lose enough mass and keep $r_\rmn{max}>r_\rmn{h}$ but also whether the orbit can effect this degree of stripping. 

The requirement to match the orbits of satellites raises issues about stochastic variations in halo accretion histories. Imposing orbit conditions requires that our subhalo-satellite matches be constrained by whether a given host halo accreted a satellite of the correct mass onto the correct orbit at the correct time. The matching rate for subhaloes and satellites becomes a function of the host halo formation history as well as the dark matter model, thus obscuring our ability to learn about the latter. Therefore, we develop a subhalo orbit -- subhalo stripping viability criterion below but apply it more loosely in our results section than our other matching criteria, in order to avoid inaccurately drawing conclusions about dark matter models that actually arise from the stochastic halo formation histories.

We stated at the start of this subsection that the degree of stripping a subhalo experiences should be commensurate with the subhalo orbit. A more precise expression is that the subhalo needs to have been exposed to a tidal field of sufficient strength to experience the stripping required to match the observed satellite mass. A definitive analysis of the tidal field requires dedicated simulations at very high resolution and is beyond the scope of this paper. Therefore, as a rough indicator we instead consider the tidal radius of the subhalo measured at its first pericentre. We can also infer a tidal radius for our stripped density profiles from where the stripped profile becomes much steeper than the unstripped version. Finally, we can compare the tidal radius inferred at first pericentre for the MW mass distribution, which we denote $r_\rmn{tid,pe}$, to the tidal radius of the stripped density profile, represented by the symbol $r_\rmn{tid,st}$, and consider the two commensurate if their ratio is less than some value.

The value of $r_\rmn{tid,pe}$ for a given subhalo relies on the subhalo mass plus the matter distribution of the host halo and its central galaxy, where the baryonic galaxy will play a significant role especially at small pericentres. We therefore follow the procedure of \citet{Lovell21}, who applied the spherically averaged present day MW mass distribution measured by \citet{Cautun20} as an input for the tidal radius equation presented in \citet{Springel08b}. For our purposes, this equation takes the form:

\begin{equation}
    r_\rmn{tid,pe} = \left[\frac{M_\rmn{sub}}{[2-d\ln M_\rmn{host}/d\ln{R_\rmn{pe}}]\times M_\rmn{host}(<R_\rmn{pe})}\right]^{1/3}\times R_\rmn{pe},
\end{equation}

\noindent where $M_\rmn{sub}$ is the total mass of the subhalo -- which we take to be $M_\rmn{200,i}$ -- $R_\rmn{pe}$ is the distance to the host centre at pericentre, and $M_\rmn{host}$ is the cumulative mass profile of the MW, for which we use the \citet{Cautun20} result. Through this method we take into account the additional tidal field due to the baryonic disc, which is not present in the simulations themselves. We apply the same mass profile at all redshifts, therefore the approximation will be less at accurate at earlier times when the MW disc is less massive than in the \citet{Cautun20} profile and thus will lead to more stripping. 

We define the tidal radius of the stripped profile, $r_\rmn{tid,st}$, as the radius at which the stripped profile has a slope of $-4$. We then compute the ratio $r_\rmn{tid,st}/r_\rmn{tid,pe}$, and accept subhaloes as a suitable match for the given satellite if $\exp(|\log(r_\rmn{tid,st}/r_\rmn{tid,pe})|)<2$. A comprehensive study would also consider the impact of subsequent pericentres; however, we note that it has been argued the later pericentres frequently occur at greater distances from the host centre than does first pericentre \citep{Santistevan23}, and therefore it is not clear that the satellites would indeed experience a stronger tidal field at later times.

\section{Stripping of individual satellites}
\label{sec:indsats}

In this section we present the subhalo population properties prior to stripping, explore the effect of stripping on subhaloes, and present the probability that subhaloes of a given mass are able to fit different classes of satellite galaxies. 

\subsection{Pre-infall subhalo circular velocity profiles}

To begin, we consider the distribution of unstripped halo circular velocity profiles, labelled $V_\rmn{c}(r)$, for the three models in the context of the observed MW satellites' $V_\rmn{h}$--$r_\rmn{h}$ values. We also introduce the $V_\rmn{max}$--$r_\rmn{max}$ relations for CDM and WDM, which \citet{Errani22} demonstrated is the boundary on the size of haloes. 

\begin{figure*}
    \centering
    \includegraphics[scale=0.35]{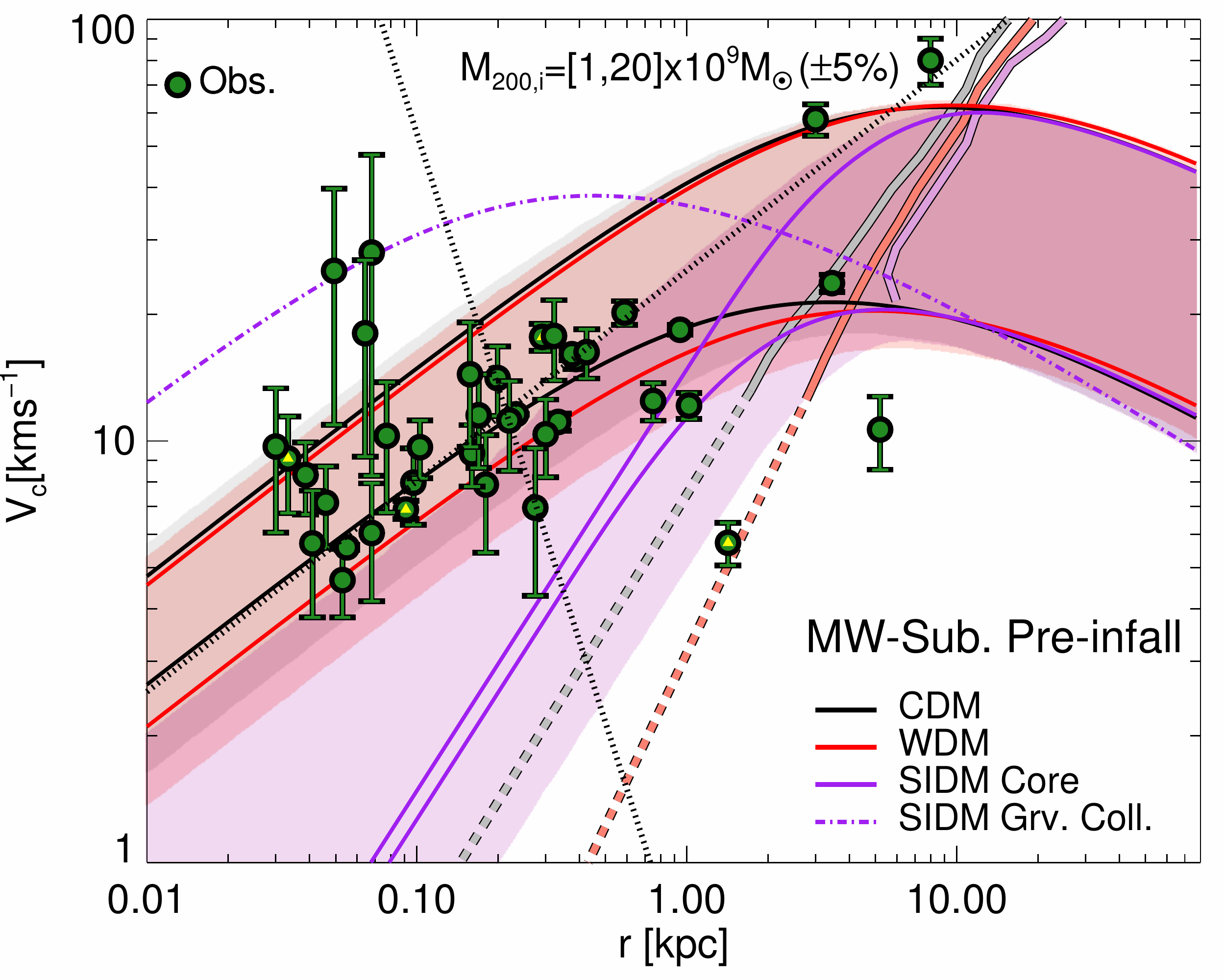}
    \includegraphics[scale=0.35]{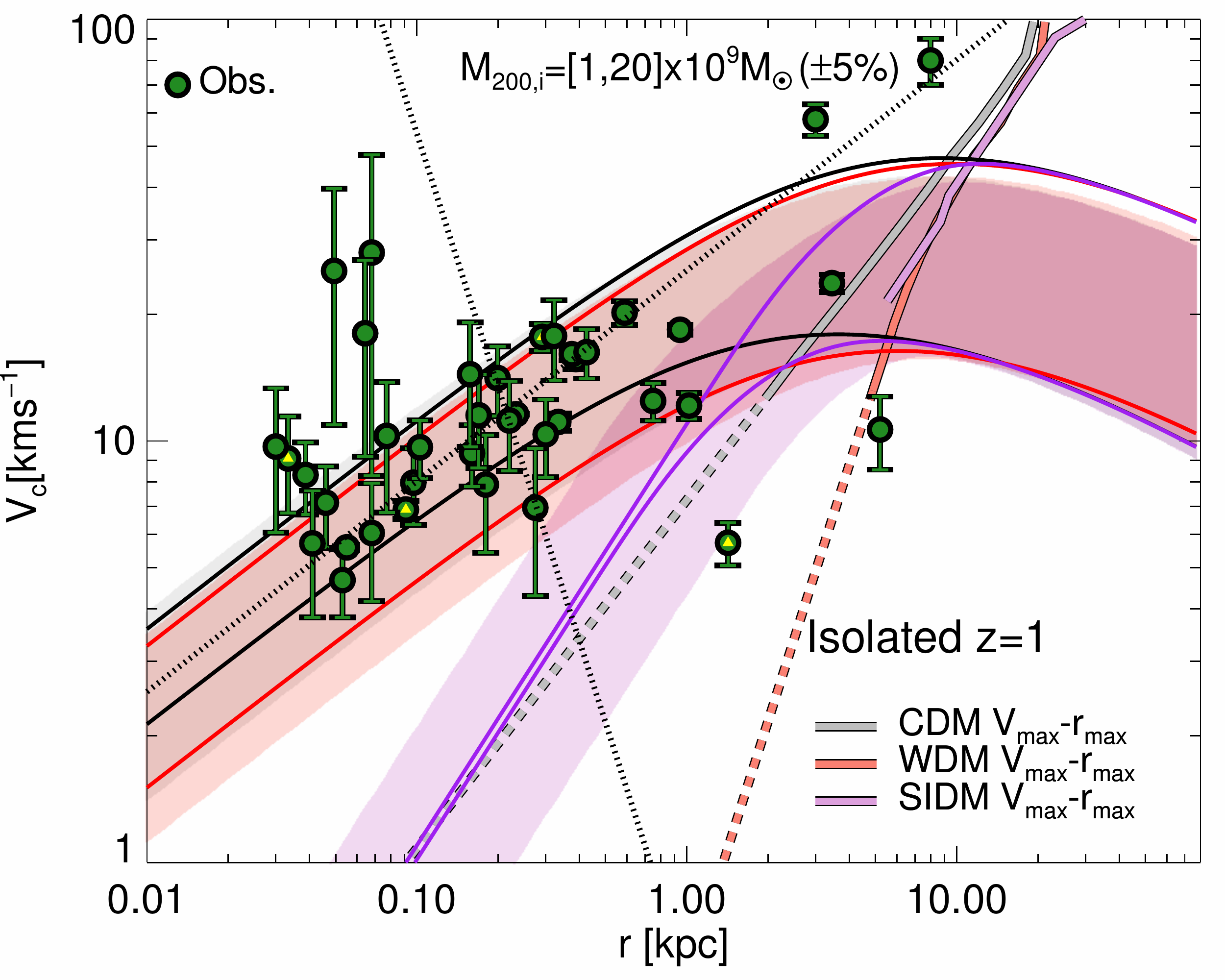}
    \caption{The range of circular velocity curves available for modelled yet-to-be-stripped haloes in the mass range $[1,20]$~$\times10^{9}$~$\msun$. We show the 95~per~cent range of the data for the three models as a series of bands: light grey for CDM, light red for WDM and light purple for SIDM. The stacked circular velocity profiles for $1\times10^{9}$~$\msun$ and $2\times10^{10}$~$\msun$ haloes are shown as black lines for CDM, red lines for WDM, and purple lines for SIDM. In the SIDM $1\times10^{9}$~$\msun$ case we exclude haloes that have undergone collapse, and instead represent these haloes with the dot--dashed purple line. The median $V_\rmn{max}$--$r_\rmn{max}$ relations for isolated haloes are shown as grey, light red and light purple thick solid lines for CDM, WDM and SIDM respectively, and for CDM and WDM are extrapolated to lower $V_\rmn{max}$ with dashed lines. We do not include an extrapolation for SIDM due to the large scatter introduced by gravothermal collapse. Observed satellite $V_\rmn{h}$--$r_\rmn{h}$ are shown as green circles. The two black dotted lines separate the satellites into high density/low density and large size/small categories. The four satellites that we take for our following case studies -- Willman~I, Draco, Carina~II and Crater~II -- are indicated with a yellow triangle inside its symbol.  We show the pre-infall set of MW-host subhaloes in the left-hand panel and the $z=1$ set of isolated haloes in the right hand panel.}
    \label{fig:satsum}
\end{figure*}

We first restrict our analysis to the more massive available subhaloes, which we define as the mass range $M_\rmn{200,i}=[1,20]\times10^{9}$~$\msun$. The ability to fit a satellite with this family of subhaloes (pre-stripping) is set by the range of circular velocities at the given $r_\rmn{h}$ whereas the stripping rate is also impacted by the shape of the profile. We can indicate the set of available $V_\rmn{c}$ across the full mass range, and thus demonstrate the mean $V_\rmn{c}$ profile shape with stacked profiles. We thus compute stacked $V_\rmn{c}$ profiles for haloes of $M_\rmn{200,i}$ within 5~per~cent of $1\times10^{9}$~$\msun$ and also within 5~per~cent of $2\times10^{10}$~$\msun$. $V_\rmn{max}$--$r_\rmn{max}$ relations are computed for haloes of $V_\rmn{max}>13$~$\kms$, and we then extrapolate these relations to lower masses with a power law. We present the results in Fig.~\ref{fig:satsum} for both the infall halo set and the $z=1$ isolated halo set in order to show the impact of accreting satellites that form across cosmic time compared to using a single snapshot. 

In the same figure, we separate our satellites into four broad classes of satellites based on their size and density, and then choose one satellite to represent each class in the remainder of our analysis. Given that the rising part of the NFW $V_\rmn{c}$ profile has a logarithmic slope of 0.5, we separate satellites into high-density and low-density categories by drawing a power law of slope 0.5 that separates the population into two subpopulations with equal numbers of satellites. We then delineate a large size and small size population by drawing a power law perpendicular to the slope of 0.5 -- that is, a slope of -2 --  and so compute the line of slope -2 that splits the population in half. We thus obtain the two dotted lines in each panel of Fig.~\ref{fig:satsum}. Finally, we select one satellite from each of the four regions: Draco for large-size-high=density, Willman~I for small-size-high-density, Carina~II for small-size-low-density, and Crater~II for large-size-low-density. 

The CDM and WDM models show similar profiles at large radii, and diverge to lower radii as one would expect for the lower WDM concentrations. The two regions start to diverge for radii $<10$~kpc at the low-mass end and to a much lesser degree at the high-mass end, with the result that the distribution of available WDM profiles is broader than for CDM. For the $M_\rmn{200,i}=2\times10^{10}$~$\msun$ haloes, the stacked profiles are the same for CDM and WDM at the per~cent level, whereas for the lower mass $M\rmn{200,i}=10^{9}$~$\msun$ haloes, the WDM profiles are suppressed at the order of 12~per~cent relative to CDM for radii $<1$~kpc despite the close agreement between the two at large radii. This property increases the available diversity in rotation curves for WDM relative to CDM: the most massive WDM subhaloes are nearly as capable at fitting high-mass satellites as CDM, whereas lower mass WDM subhaloes have a significant edge in fitting the less massive satellites. The full scatter in WDM allows for still lower mass satellites to be matched.
The subsequent challenge for WDM is whether it can produce enough high-mass subhaloes in MW-analogue hosts to match the number of satellites that require high densities, which CDM can readily achieve through dense, lower mass subhaloes; we will address this question in Section~\ref{sec:satpop}. 

The change in concentration is reflected by the divergence in the $V_\rmn{max}$--$r_\rmn{max}$ relations, where the median $r_\rmn{max}$ differs by $\sim20$~per~cent at $V_\rmn{max}\sim30$~$\kms$ but by a factor of 2 at 13~$\kms$. Crucially, the WDM  $V_\rmn{max}$--$r_\rmn{max}$ extrapolation passes to the right of the Crater~II satellite and may plausibly ease the tension deduced for CDM by \citet{Errani22}; however, Antlia~II remains to the right of the WDM $V_\rmn{max}$--$r_\rmn{max}$ relation and therefore a change in the dark matter model alone may not be sufficient to resolve this question.

The $M_\rmn{200,i}=2\times10^{10}$~$\msun$ SIDM haloes show large cores, by construction, which decreases the degree of stripping that will be required to match the low-density satellites. Paradoxically, the $1\times10^{9}$~$\msun$ haloes have {\em higher} stacked densities, which we attribute to early forming, early accreting subhaloes that have a small scale radius, and also includes some subhaloes that have undergone gravothermal collapse. However, this increased density is not high enough to match most of the satellites. Finally, we note that SIDM simulations present a lot of parameter space below the observational data. This may either indicate the existence of further Crater~II-like objects that are yet to be discovered or instead of a population of dark, massive subhaloes, notwithstanding the limitations of the modelling apparatus.

There is a significant difference between the MW-analogue satellite-derived sample and the $z=1$ sample, with the latter returning lower amplitude circular velocity curves in general and a sharper divergence between CDM and WDM in particular. Some degree of difference is expected due to the different formation times, with some subhaloes in the infall sample having formed earlier than $z=1$ dwarf haloes and thus obtained higher densities as discussed in the context of SIDM haloes. This difference is crucial for the massive, low-density satellites Crater~II and Antlia~II, where the transition from the infall to the $z=1$ sample removes the tension for Crater~II and sharply reduces the tension for Antlia~II. On the other hand, even the most massive CDM subhaloes predict lower densities than are reported for the small, high-density satellites, thus making the argument that these satellites must have formed early and been accreted early. Finally, the restriction to one redshift enforces similar core sizes on both high-mass and low-mass SIDM haloes, and strongly restricts the ability to fit most of the observed satellites without gravothermal collapse.

\subsection{Satellite properties post-stripping}

We now turn to the inferred properties of satellites after stripping. We set the following criteria to determine whether a given subhalo is a viable host for each satellite:

\begin{itemize}
    \item The infall circular velocity at $r_\rmn{h}$ is greater than $V_\rmn{h}$ as stated above.
    \item The subhalo infall mass is in agreement with the \citet{Behroozi13} expectation for the stellar mass--halo mass relation.
    \item The stripped $r_\rmn{max}$ is larger than the satellite $r_\rmn{h}$. 
    \item The accretion lookback time is less than the merger time (see eqn.\ref{eqn:tmerge}).
    \item The first pericentre impact parameter is $>10$~kpc from the host centre.
    \item The ratio of tidal radii is less than a factor of two, or $\exp(|\log(r_\rmn{tid,st}/r_\rmn{tid,pe})|)<2$.
\end{itemize}

\begin{figure*}
    \centering
    \includegraphics[scale=0.32]{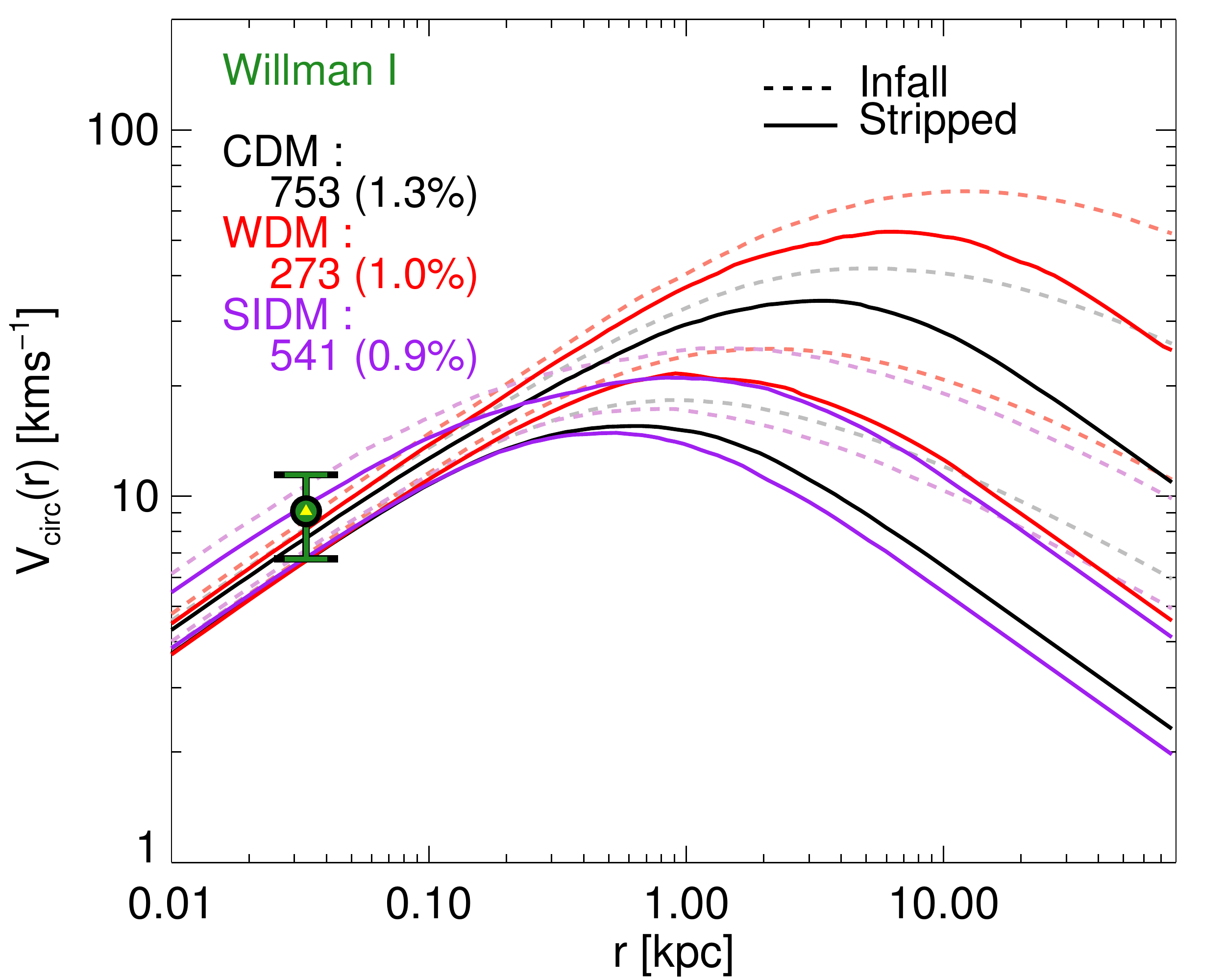}
    \includegraphics[scale=0.32]{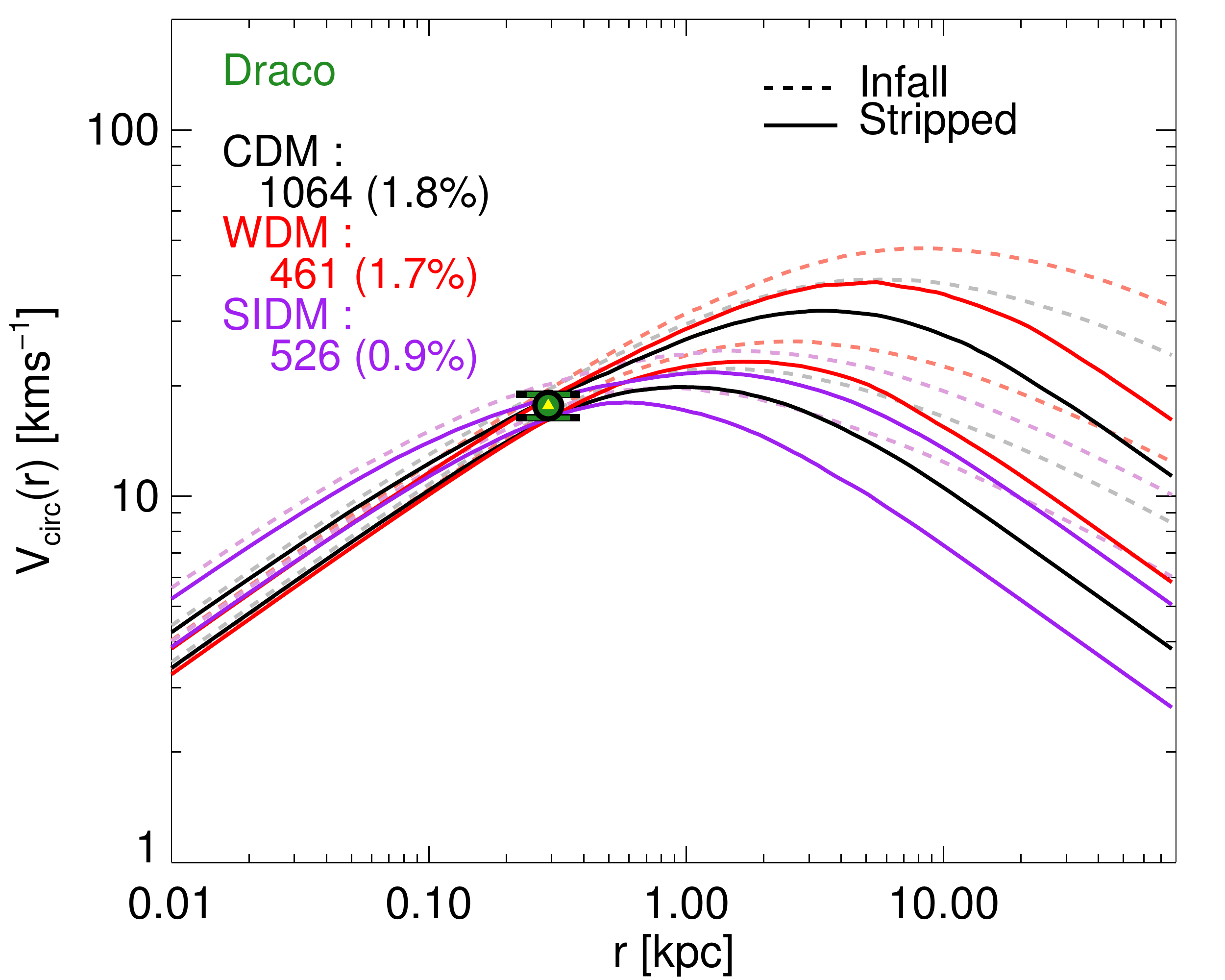} \\
    \includegraphics[scale=0.32]{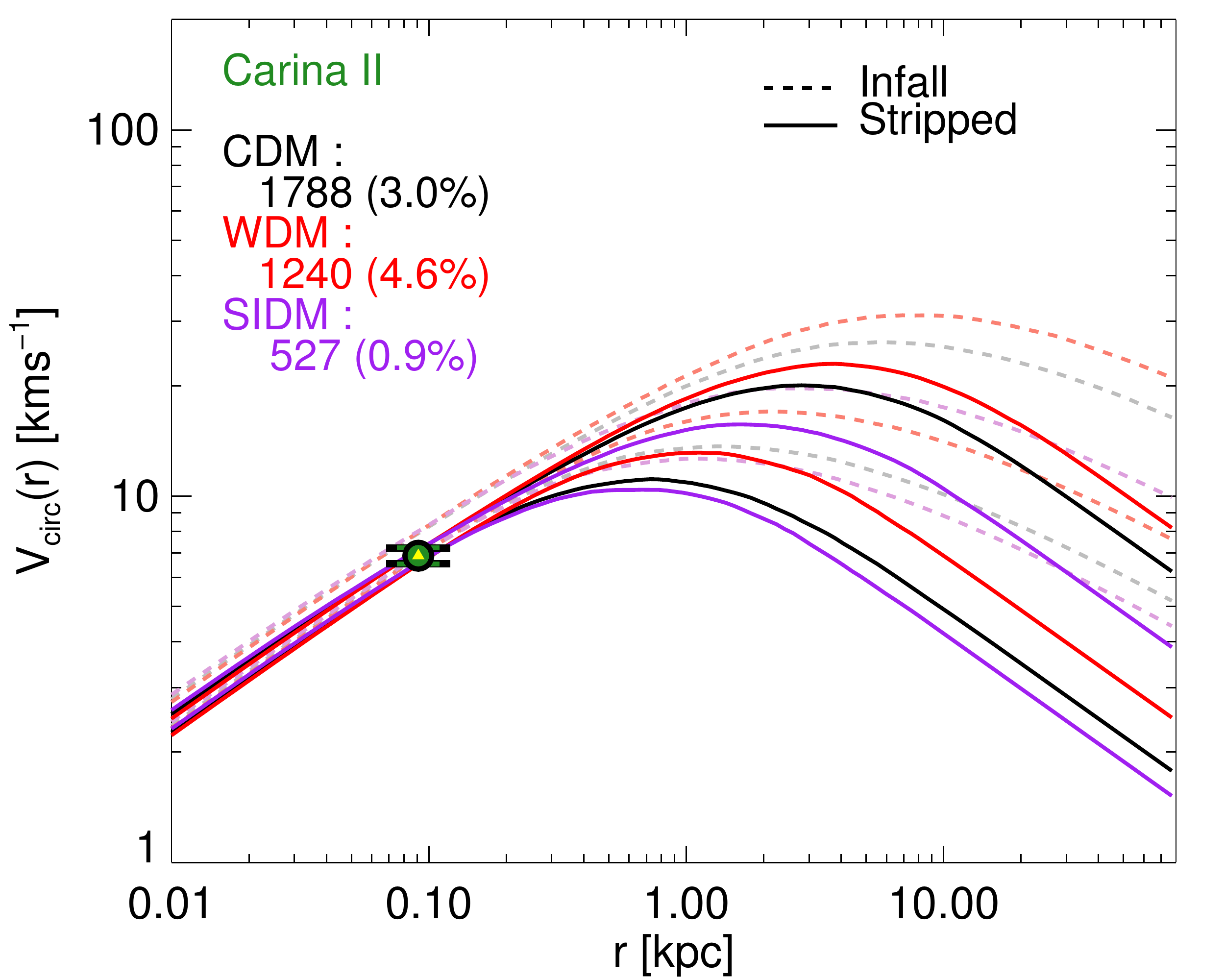} 
    \includegraphics[scale=0.32]{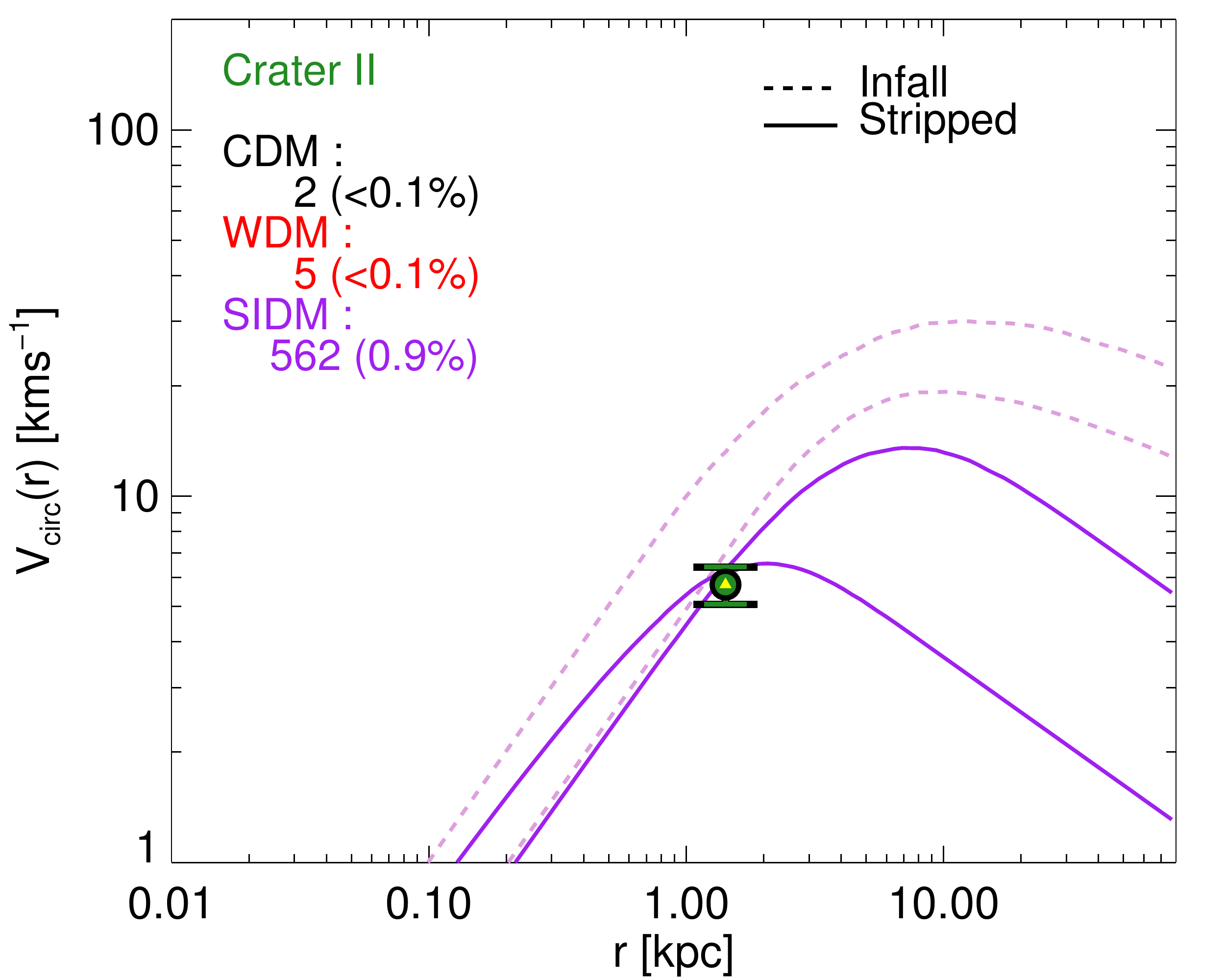}
    \caption{Circular velocity curve fits for four satellites: Willman~I (top-left panel), Draco (top right), Carina~II (bottom left) and Crater~II (bottom right). CDM data are shown in black, WDM in red, and SIDM in purple. The $V_\rmn{h}$--$r_\rmn{h}$ for each satellite is shown as a green and yellow circle, with the error bars showing 68~per~cent uncertainty. Each pair of simulation lines delineates the 68~per~cent region of the circular velocity profiles: solid lines show the stripped curves and faded, dashed lines show the infall curves. We only include curves for models in which more than five subhaloes with $M_\rmn{200,i}>10^{8}$~$\msun$ can fit the given MW satellite successfully. The percentage and total number of haloes with infall masses $M_\rmn{200,i}>10^{8}$~$\msun$ that obtain a match are shown in the figure legend.}
    \label{fig:ranges}
\end{figure*}

For a first look at the stripped satellite results, we take the subhaloes that match the above criteria and compute the range of their circular velocity curves. This range will in practice be a function of both the concentration of the subhaloes and their abundance: for example, if a given satellite is well fitted by low-mass ($<1\times10^{9}$~$\msun$) CDM subhaloes, the computed range will exclude the small number of higher mass subhaloes that also provide a good fit purely because there are so many more low-mass subhaloes in CDM. We compute the 16-84~per~cent of the range of $V_\rmn{c}(r)$ values at each radius, both for the unstripped haloes and their stripped versions. We consider the four satellites that we highlighted in the previous section -- Draco, Willman~I, Carina~II, and Crater~II -- to represent different parts of the observed satellite size-density space. The resulting circular velocity ranges for CDM, WDM and SIDM haloes are presented in Fig.~\ref{fig:ranges}.       

The three models react very differently to these four satellites. Carina~II is the satellite that CDM and WDM match best, all matching at between 3.0~per~cent and 4.6~per~cent of the available subhaloes respectively; SIDM lags with only 0.9~per~cent matches. The ranges of suitable halo matches are infall $V_\rmn{max}=[11,23]$~$\kms$ for CDM, $V_\rmn{max}=[14,30]$~$\kms$ for WDM and $V_\rmn{max}=[9,20]$~$\kms$ for SIDM. The three models require very little the same amount of stripping at $r_\rmn{h}$,  which we will explore below. The stripping picture at large radii is more complicated, although ultimately the largest host subhaloes post-stripping are in WDM, followed by SIDM and with CDM as the lowest host mass subhaloes of the three.

The difference between WDM and CDM is similar for massive Draco hosts. The primary difference between the two is that CDM can fit Draco in dense subhaloes of infall $V_\rmn{max}$ around 10~$\kms$ lower than for WDM. Much more strongly affected is SIDM, where the large cores -- which are unavoidably generated in all massive haloes -- undershoot the observed $V_\rmn{h}$ and so fail to fit these satellites without including gravothermal collapse \citep{Zavala19a}. Therefore, subhaloes that fit 
 Draco have $V_\rmn{max}<20$~$\kms$ for the vd100 model and thus Draco becomes a fundamentally very different object to lower density satellites such as Sextans that instead preferred a core. The degree of collapse in Draco is such that we predict the SIDM haloes have enclosed densities higher than CDM at radii of 100~pc.

The situation for the dense Willman~I galaxy is very similar to that of Draco, including very similar matching percentages for CDM and WDM ($\sim1.3$~per~cent and 1.0~per~cent respectively),  with massive WDM hosts ($V_\rmn{max}=[20,60]$~$\kms$), a narrower range of less massive CDM hosts ($V_\rmn{max}=[12,33]$~$\kms$), and SIDM hosts restricted to haloes with $V_\rmn{max}<20$~$\kms$ when including gravothermal collapse. The amount of stripping is similar to Carina~II, and CDM and WDM expect a `true' value of $V_\rmn{h}$ that is smaller than the reported mean value. 

Finally, Crater~II shows very different results, favouring more massive, cored SIDM haloes (1.1~per~cent of haloes). CDM and WDM have only a handful of haloes that can fit Crater II, two and five, respectively, and thus their results are not shown. The stripping rate is high, even for SIDM haloes. The maximum available $V_\rmn{max}$ for SIDM is only 25~$\kms$ despite its large size. We attribute this result to the fact that the vd100 model uses haloes derived from the CDM mass--concentration relation: high-mass haloes are stripped along this relation as discussed by \citet{Errani21}, and so the presence of a core that exists only well within the halo scale radius $r_\rmn{s}$ is unable to address the problem raised in CDM. WDM has the benefit of increasing the scale radius, but it is unclear whether the effect is large enough for a massive WDM to host Crater~II. 

We will expand this analysis to all satellites by considering two of our key criteria for subhalo--satellite matches. First, is the stripped $r_\rmn{max,st}$ bigger than $r_\rmn{h}$, and second, can this stripping be supported by the satellite orbit? We consider these options across dark matter models and halo masses, and demonstrate which combination of models and halo masses fit which satellites using the algorithm discussed below.

\begin{figure*}
    \centering
     \includegraphics[scale=0.68]{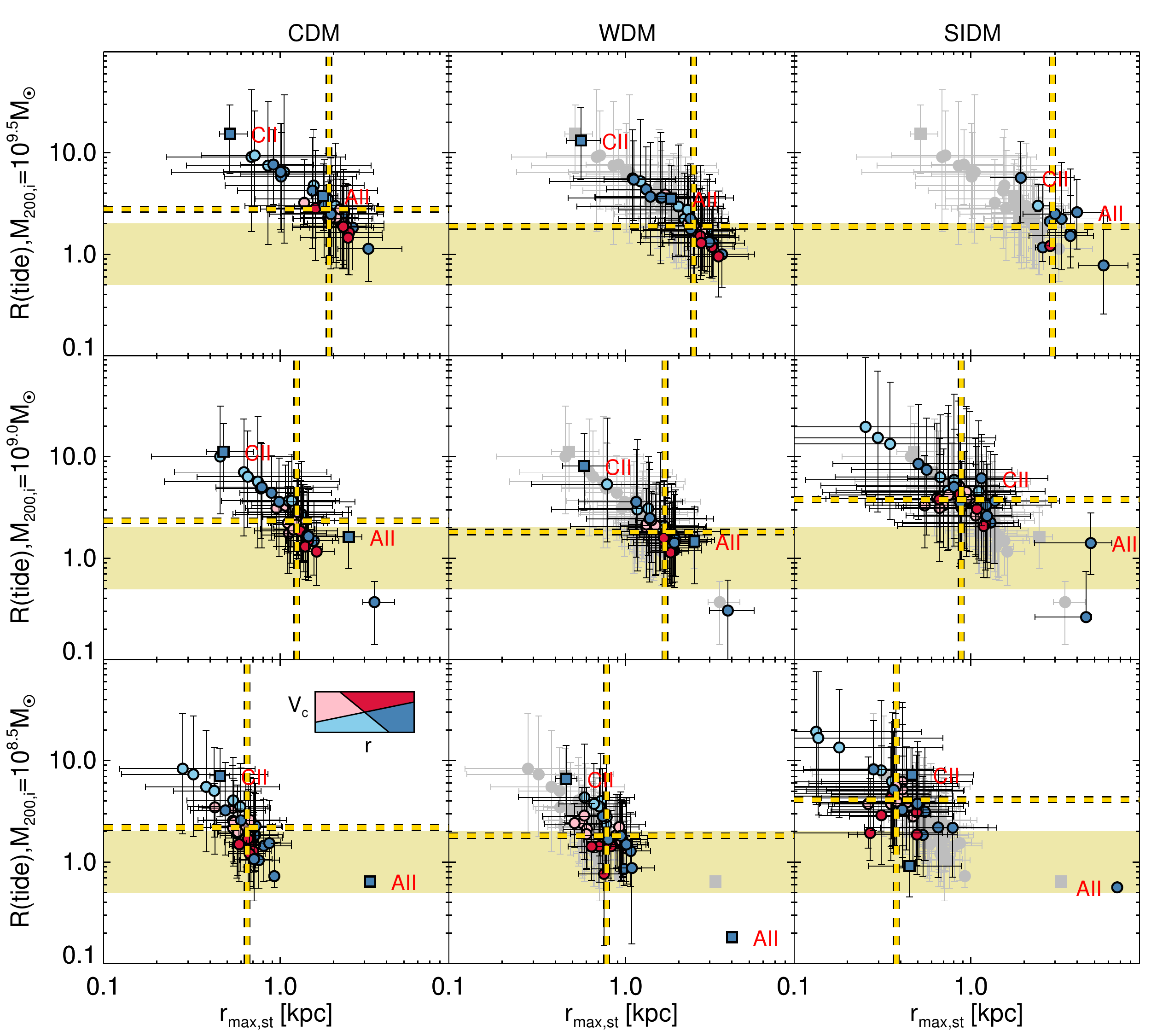}
     \caption{The distribution of tidal radius ratios $R=(r_\rmn{tid,pe}/r_\rmn{tid,st})$ and stripped $r_\rmn{max,st}$ for different dark matter models and different halo masses.  CDM results are shown in the left-hand panels, WDM results in the middle panels, and SIDM results in the right-hand panels. The CDM data points are also replicated in the other two columns as grey symbols for comparison. The rows correspond to different infall halo mass: $10^{9.5}$~$\msun$ (top panel), $10^{9.0}$ (middle panel), and $10^{8.5}$~$\msun$ (bottom panel).  Each symbol represents the median data values for the parameters that an observed MW satellite would have for a given dark matter model fit. Satellites for which fewer than 16~per~cent of hosts return $r_\rmn{max,st}>r_\rmn{h}$ are shown as squares, and the remainder as circles. The symbols' colours encode the region of $V_\rmn{h}$-$r_\rmn{h}$ space in Fig.~\ref{fig:satsum} that each satellite belongs to, as indicated by the inset in the bottom-left panel. Error bars indicate the 68~per~cent range of the data. The beige region indicates the tidal radius ratio criterion that we choose in this paper. The vertical and horizontal dashed lines indicate the median $r_\rmn{max,st}$ and tidal radius ratio respectively across their median values over all satellites. The farther the median value is (dashed yellow lines) from the beige regions, the more discrepant the model--halo mass combination is in regards to a plausible stripping scenario for each satellite. Finally, Crater~II (CII) and Antlia~II (AII) are highlighted with red letters.}
    \label{fig:radstVtidr}
\end{figure*}

We select three halo masses for our analysis: $M_\rmn{200,i}=10^{8.5},10^{9.0}$ and  $10^{9.5}$~$\msun$. For each of these three masses, we select haloes for which $M_\rmn{200,i}$ is within 10~per~cent of the mass in question. We then compute the distribution of $r_\rmn{max,st}$ and tidal radius ratio $R=r_\rmn{tid,pe}/r_\rmn{tid,st}$ for these halo--satellite combinations and present the results in Fig.~\ref{fig:radstVtidr}. In the same figure we indicate the degree to which the $r_\rmn{max,st}$ criterion is met through the symbol shape: satellites for which fewer than 16~per~cent of satellites obtain $r_\rmn{max,st}>r_\rmn{h}$ are shown as squares, while the remainder are shown with circles.   

We begin with the analysis of the $r_\rmn{max,st}>r_\rmn{h}$ criterion. CDM of mass $M_\rmn{200,i}=10^{9.5}$~$\msun$ can fit satellites under almost all circumstances with $r_\rmn{max,st}>1$~kpc, and with a median of 2~kpc can comfortably host almost all satellites. The two conspicuous exceptions are Crater~II and Antlia~II, for which $r_\rmn{h}$ is 1.3 and 5.2~kpc respectively, yet the model prefers 500~pc for Crater~II and 2~kpc for Antlia~II. The change to lower mass haloes shifts the $r_\rmn{max,st}$ distributions progressively to smaller radii; the other satellites' radii remain viable while Crater~II and Antlia~II still struggle. 

The tidal radius requirement is the more demanding criterion, with many $10^{9.5}$~$\msun$ CDM haloes requiring a stripping tidal radius that is between a factor of 2 and a factor of 10 smaller than the subhalo orbit can support. The points shift downwards slightly for smaller host masses, such that the median host tidal radius ratio drops from 2.3 at $10^{9.5}$~$\msun$ to 2.1 at $10^{8.5}$~$\msun$. In all cases the satellites that are overdense relative to the median as defined in Fig.~\ref{fig:satsum} (coloured red and pink) have lower tidal radius ratios than then underdense counterparts (dark blue and light blue), since less stripping is required to match these subhaloes and so they can be accommodated on a larger orbit with a weaker tidal field. 

The WDM results exhibit the same trends as CDM but experience a significant shift thanks to the change in mass--concentration relation. Starting once again with the $10^{9.5}$~$\msun$ haloes, the median required $r_\rmn{max,st}$ is $\sim$2.3~kpc compared to 1.9~kpc for CDM. The shift is demonstrated by comparing the WDM points to the greyed-out CDM points. For Crater~II this indicates a shift of up to 100~pc in $r_\rmn{max,st}$: not enough to solve the discrepancy on its own but potentially a useful contribution when combined with supernova feedback. That the $r_\rmn{max,st}$ is larger in WDM than in CDM means that less stripping is required, which is obtained on a larger orbit, as encoded by the tidal radius. Given orbits with larger pericentres, the percentage of subhaloes that meet the ratio factor of two criterion for a match increases from CDM to WDM; the mean tidal ratio is then 1.9 for $10^{9.5}$~$\msun$ haloes and 1.8 for both $10^{9.0}$~$\msun$ and $10^{8.5}$~$\msun$ haloes.

SIDM instead returns very different results. The presence of cores in $10^{9.5}$~$\msun$ haloes enables low-density subhaloes to be fit with small amounts of stripping, and increases the matching ability of Crater~II and Antlia~II substantially. However, the cores are so large that the infall $V_\rmn{c}(r_\rmn{h})<V_\rmn{h}$, and the lack of gravothermal collapse in these massive haloes makes it impossible to even attempt a match for most satellites and thus few points can be plotted for $10^{9.5}$~$\msun$. The transition to $10^{9.0}$~$\msun$ haloes brings gravothermal collapse, and thus many satellites achieve sufficient density to accomplish a match. The wide range of possible profiles inferred from the AqA3-vd100 simulation results in very large distributions of both $r_\rmn{max,st}$ and tidal radius ratios. Most satellites can therefore easily obtain a match as defined by our criteria, although many collapsed subhaloes will be so overdense that they have to undergo considerable stripping to match $V_\rmn{h}$ that the agreement with the tidal radius ratio criterion is worse than for CDM, with a median of 2.9 for $10^{9.0}$~$\msun$ and 3.0 for $10^{8.5}$~$\msun$. It is also the case that the collapsed SIDM haloes are much smaller than their CDM and WDM counterparts, to the degree that the median $r_\rmn{max,st}$ at $10^{8.5}$~$\msun$ is only 420~pc and thus too small to fit some satellites. 

In conclusion, we have shown that it is possible to find subhaloes large enough to host most satellites. The lower density satellites need more stripping, and thus require smaller orbits to achieve this stripping. The agreement is best for low-mass WDM satellites as a result of the WDM mass--concentration relation, and cores are preferred for some massive satellites. Given the importance of the orbits, future work will major on more precise measurements of the tidal field across the orbit, including subsequent pericentric passages.

\begin{figure*}
    \centering
    \includegraphics[scale=0.7]{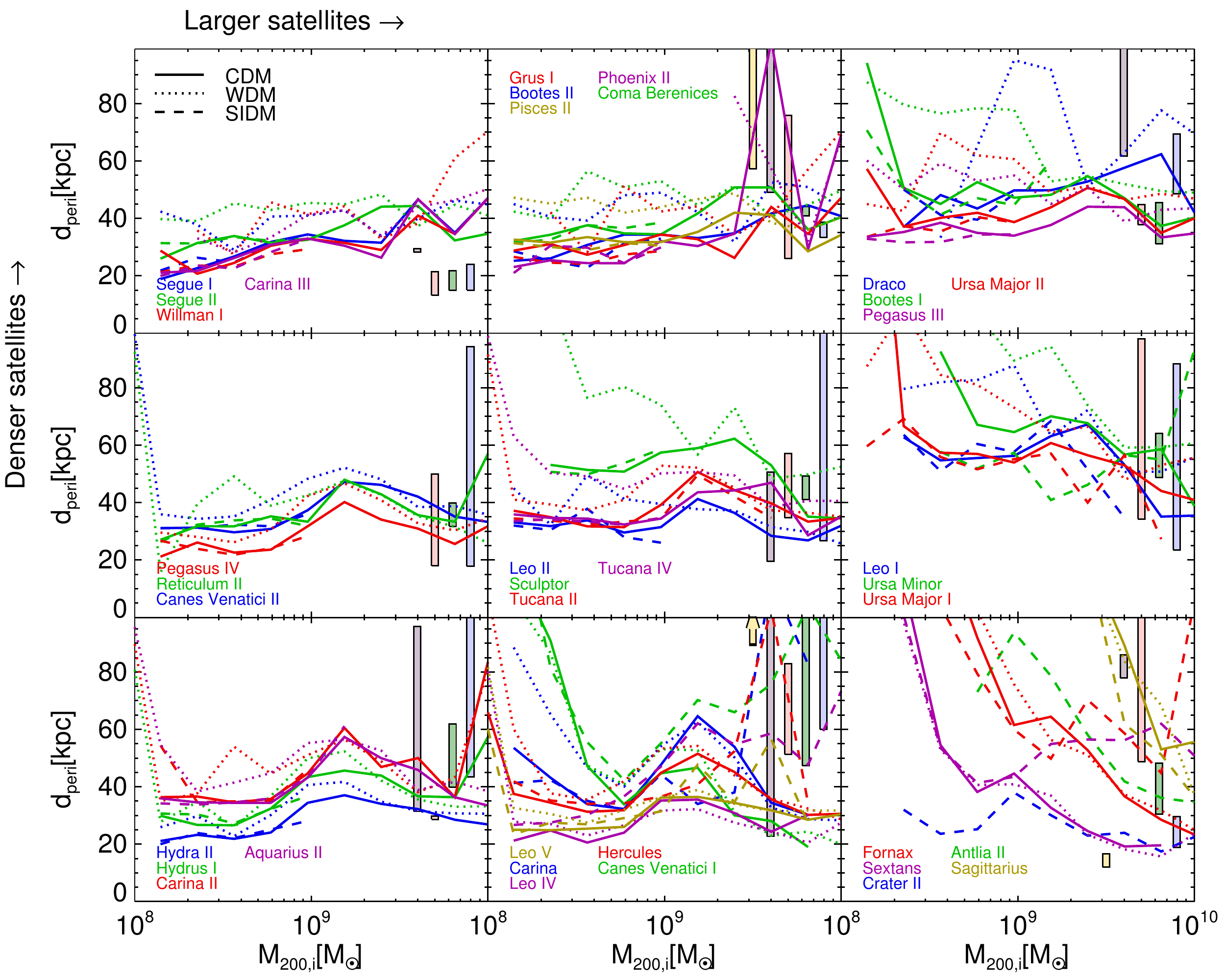}
    \caption{Median pericentre--$M_\rmn{200,i}$ relations for 37 satellites in the three dark matter models, using subhaloes that satisfy the criteria for matches to observed satellites. Each panel contains between three and five satellites. The densest satellites are indicated in the top row and the least dense in the bottom row. Columns are organized by satellite size, with smaller satellites on the left and larger satellites on the right. Solid lines indicate CDM results, dotted WDM and dashed SIDM. The relationship between satellite and colour is indicated in each panel's figure legend. No CDM or WDM data are presented for Crater~II and Antlia~II. The 68~per~cent regions for each satellite's observed pericentre are shown as rectangles with colours that are faded versions of the corresponding satellite colours; the location on the $x$-axis is arbitrary. The reported Leo~V pericentre lower bound is $>100$~kpc and therefore is replaced by an arrow.}
    \label{fig:peri}
\end{figure*}

We now highlight the impact of the pericentres on our matches between simulated subhaloes and observed satellites. For each satellite except the two Magellanic Clouds, we select the model subhaloes that meet our matching criteria for that satellite and compute the median pericentre as a function of the subhalo $M_\rmn{200,i}$. We then present the results for the three models in Fig.~\ref{fig:peri} alongside the satellite pericentres inferred from observations. We also explore the relationship between pericentre on the one hand and satellite sizes and mass on the other by so organizing these 37 satellites into nine separate panels. In the same manner that we used for the dotted lines in Fig.~\ref{fig:satsum}, we compute the power law amplitudes in $V_\rmn{c}$--$r$ of logarithmic gradient 0.5 that separate the satellite population into three bins of relative density, each hosting approximately equal numbers of satellites, and thus giving us a high-density set, an intermediate density set, and a low-density set. Within each set, we then rank the satellites by $r_\rmn{h}$, and generate three further subsets that contain roughly equal numbers of satellites in small size, intermediate size, and large size, for a total of nine subsets. Each subset is represented by a panel of Fig.~\ref{fig:peri}: satellite density decreases from top to bottom and satellite size increases from left to right. 

Subhaloes with $M_\rmn{200,i}\lsim2\times10^{9}$~$\msun$ consistently return a pericentre that is larger in WDM than in CDM. For example, in the small-size-high-density panel (top-left) the pericentres for WDM are 40~kpc across masses compared to 30~kpc for CDM. The difference is larger still for the large-size-high-density selection (top-right), where WDM subhaloes $<10^{9}$~$\msun$ expect pericentres $>50$~kpc: the low concentrations of these haloes allow for very little stripping before $V_\rmn{c}(r_\rmn{h})<V_\rmn{h}$ and so small pericentres are excluded. The low-density satellites (bottom row) return larger pericentres in CDM and WDM, especially in the large size category where low-mass subhaloes cannot undergo much stripping without violating the $r_\rmn{h}<R_\rmn{max}$ condition. 

It is in the large-size-low-density regime (bottom right) that the impact of SIDM cores is apparent, with pericentres for Sextans and Fornax at $5\times10^{9}$~$\msun$ that are $\sim60$~kpc in order to avoid over-stripping; this is compared to 20 and 35~kpc for these respective satellites in both CDM and WDM. A similar phenomenon occurs for the intermediate size -- low-density satellites, where the SIDM pericentres at $5\times10^{9}$~$\msun$ are $>45$~kpc but the CDM/WDM equivalents are $<45$~kpc. Elsewhere, most satellites can only be fit with gravothermally collapse subhaloes in the SIDM model. These subhaloes are so dense that the pericentres are typically required to be a few kpc smaller than for CDM in order to obtain enough stripping, and by construction are restricted to haloes with $M\rmn{200,i}<10^{9}$~$\msun$. 

The question of which model's pericentre distributions best match the observations is somewhat complicated. The small, high-density satellites require pericentres $<40$~kpc and are therefore better fit by CDM and gravothermally collapsed SIDM subhaloes than by WDM. On the other hand, intermediate size galaxies typically prefer pericentres $>40$~kpc and are therefore easiest to match with WDM. SIDM offers a paradoxical picture where large-pericentre-low-density haloes are matched well with cored haloes of $>10^{9}$~$\msun$ and small-pericentre-high-density haloes work with lower mass gravothermal-collapse haloes, thus supporting SIDM models that report stripping enhanced collapse \citep{Nishikawa20}, but present poor options for large-pericentre-high-density haloes, e.g. Pisces~II. In this case, a cored subhalo fails the  $V_\rmn{c}(r_\rmn{h})>V_\rmn{h}$ criterion but a gravothermal-collapse halo is so dense that it requires the stripping of a small orbit to reduce the density. Further progress in this area will require a greater understanding of the MW gravitational potential, which leads to large systematic uncertainties in these orbits (see \citealp{Collins17} for a discussion in the context of Leo~V).

\begin{figure*}
    \centering
    \includegraphics[scale=0.35]{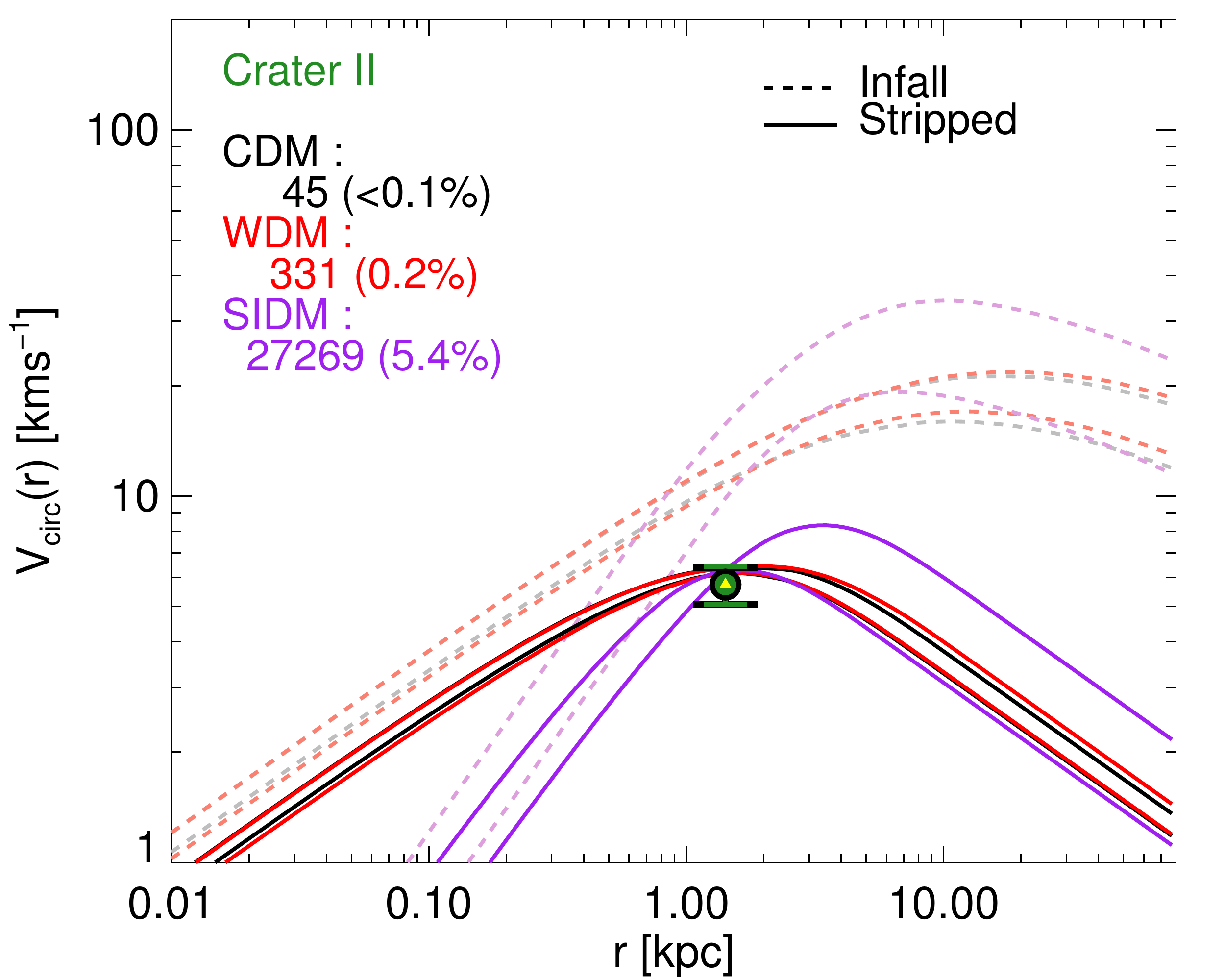}
    \includegraphics[scale=0.35]{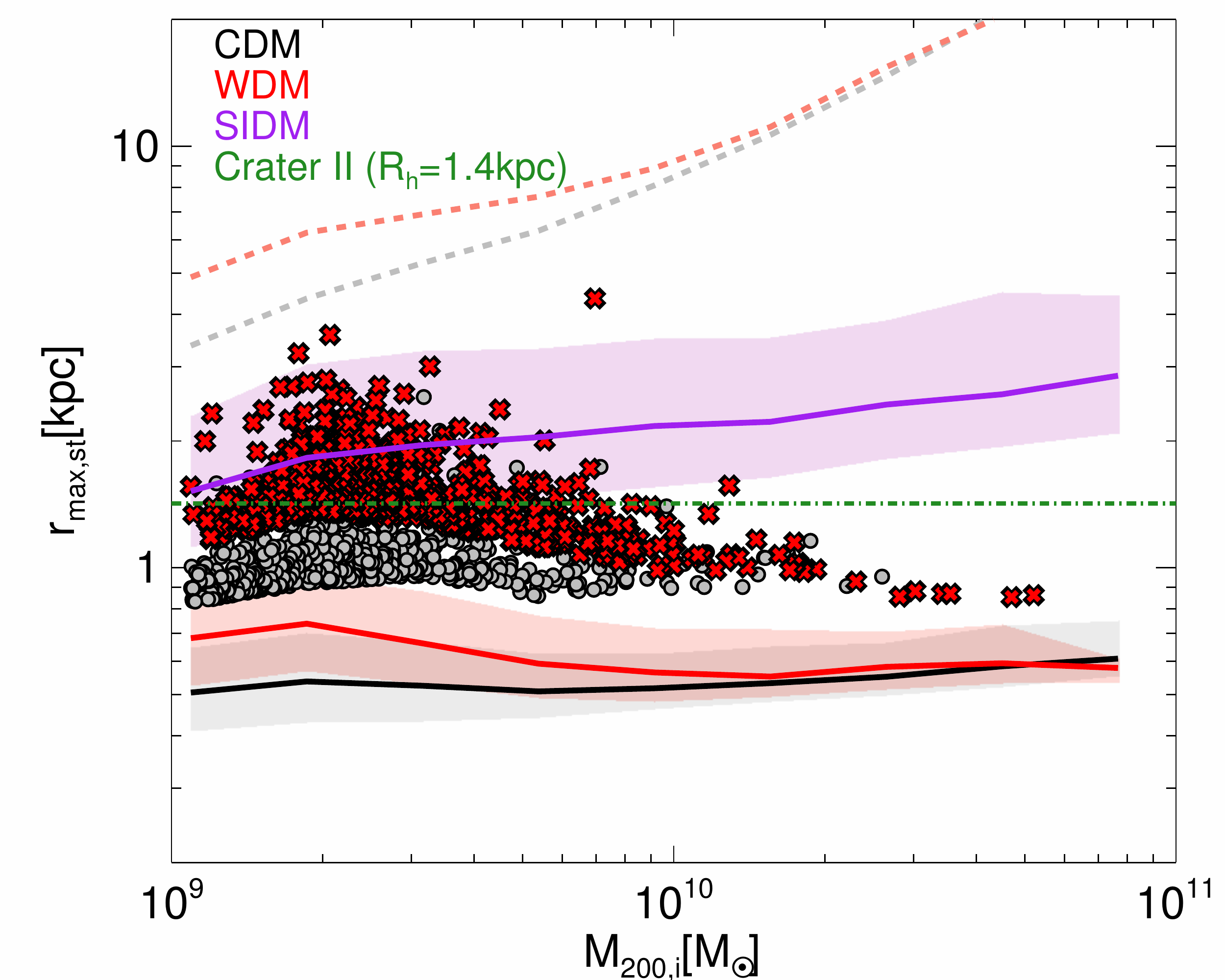}
     \includegraphics[scale=0.35]{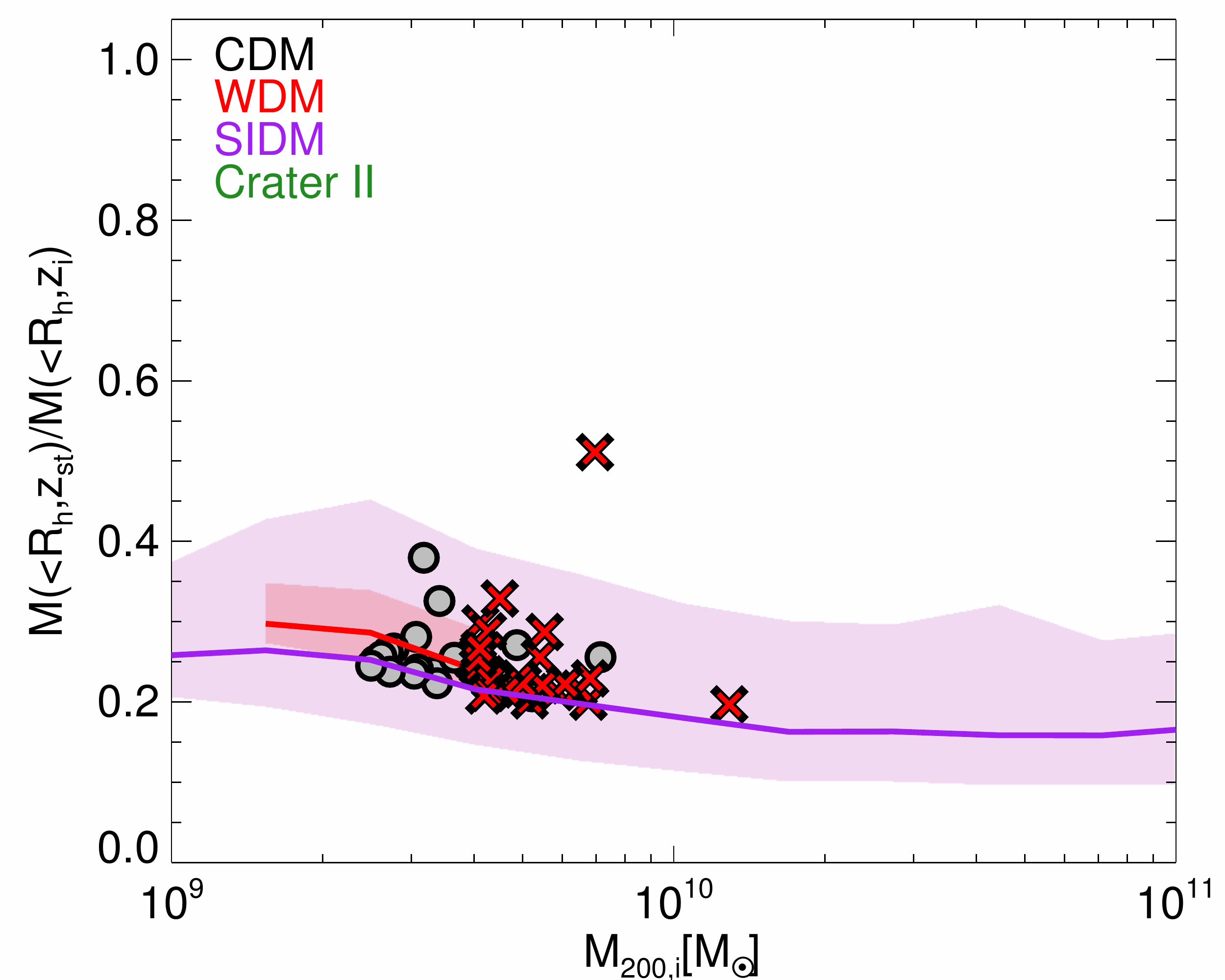}
    \includegraphics[scale=0.35]{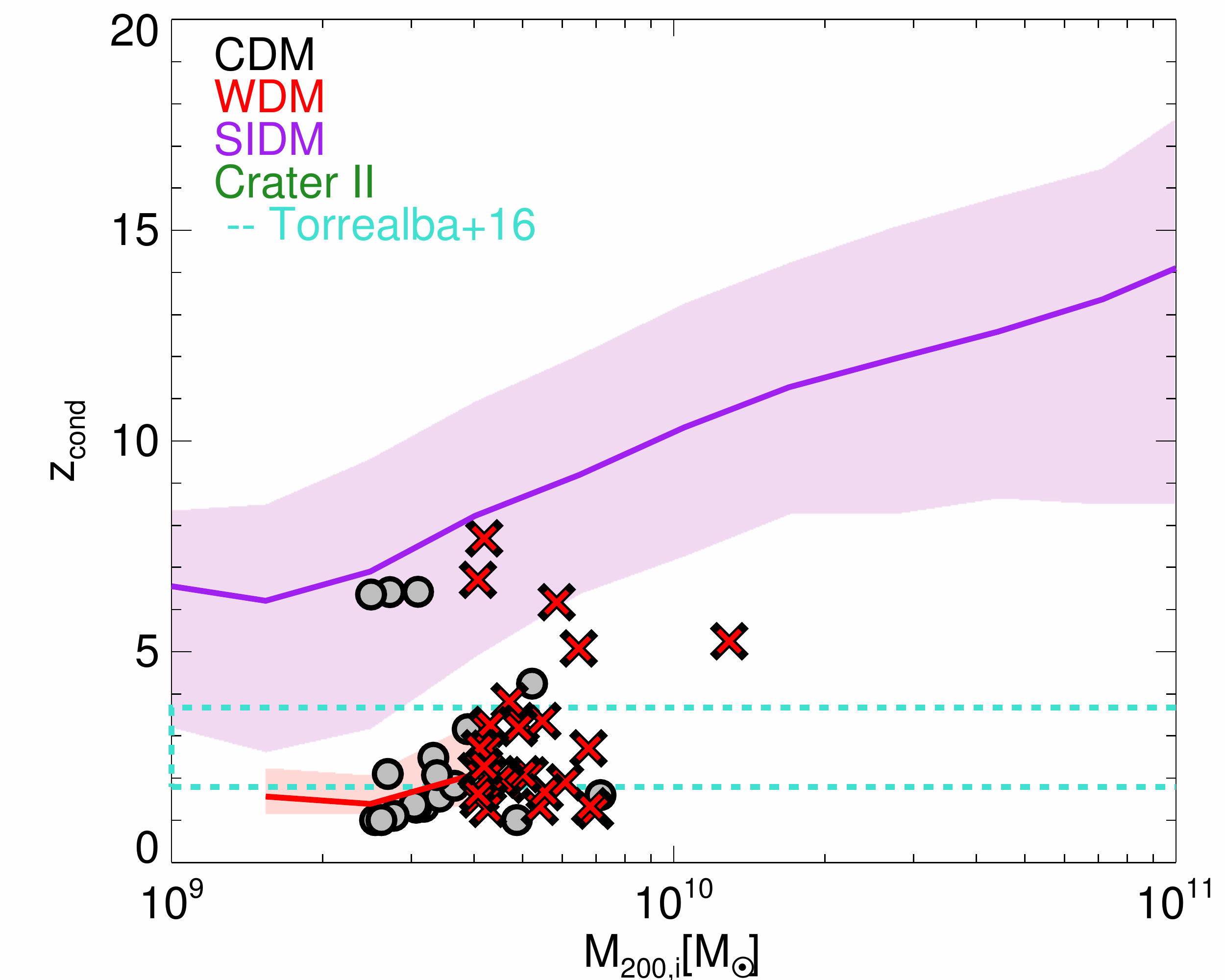}
    \caption{The circular velocity ranges (top-left panel, as described in Fig.~\ref{fig:ranges}), then stripped $r_\rmn{max}$ (top-right panel), stripped mass fraction within $r_\rmn{h}$ (bottom-left panel) and condensation redshift (bottom-right panel) as a function of $M\rmn{200,i}$ for Crater~II computed using haloes from the dwarf $z=1$ sample. In the last three panels, median CDM, WDM, and SIDM relations are shown with black, red, and purple lines respectively. Shaded regions indicate 68~per~cent of the respective distributions. In the top-right panel we include all haloes in the sample, and the largest 5~per~cent of satellites in each CDM and WDM bin are plotted as circles and crosses respectively. The light red and grey lines in this plot indicate the unstripped $r_\rmn{max,st}-M_\rmn{200,i}$ relations for CDM and WDM; the horizontal dot--dashed line indicates the measured $r_\rmn{h}$ of Crater~II. In the bottom two panels we plot data for haloes that match the $r_\rmn{max,st}>r_\rmn{h}$ criterion. For bins that contain fewer than 15 objects we plot individual CDM haloes as circles and WDM haloes as crosses; there are sufficiently few haloes in CDM that we only plot circles and not a line. Finally, the region in the $z_\rmn{cond}$ plot that matches the age of the Crater~II stellar population reported in \citet{Torrealba16} is delineated by cyan dashed lines.}
    \label{fig:z0sam}
\end{figure*}

We end this section by returning to the $z=1$ isolated dwarf population. We showed in Fig.~\ref{fig:satsum} that this selection had lower densities than the MW subhalo population at large, especially for WDM, and could therefore provide a resolution to the challenge of matching the low densities of Crater~II. We therefore repeat the estimated circular velocity ranges and stripped $r_\rmn{max}$ values for these subhaloes, and present the results in Fig.~\ref{fig:z0sam}.

The switch from the MW-analogue subhalo selection to the $z=1$ dwarf selection leads to a significant difference in the range of circular velocity profiles that can be matched to Crater~II. The matter distribution of Crater~II can be matched by stripping CDM and WDM dwarfs of infall $V_\rmn{max}=[13,20]$~$\kms$, which was not possible with the previous selection. The post-stripping profiles are also very similar in the two models. However, the number of CDM haloes that possess the required concentration is much lower than in WDM, at 45 to 331. The range of SIDM haloes that can fit Crater~II also increases for our $z=1$ halo selection, and can admit haloes as massive as $V_\rmn{max}=30$~$\kms$.  

We illustrate the Crater~II fitting potential for this population in quantitative terms with the relation between stripped $r_\rmn{max,st}$ and $M_\rmn{200,i}$. The vast majority of both CDM and WDM $z=1$ haloes return stripped $r_\rmn{max,st}<r_\rmn{h}$ at all masses. The two models diverge for $M_\rmn{200,i}<10^{10}$~$\msun$, reflecting the difference in the unstripped $r_\rmn{max,i}$--$M_\rmn{200,i}$ relations: here the distribution of WDM $r_\rmn{max,st}$ is approximately 30~per~cent higher than for CDM. This shift marks a significant increase for the number of WDM haloes with $r_\rmn{max,st}>r_\rmn{h}$, up to 2.5~per~cent of haloes at $2\times10^{9}$~$\msun$. Finally, a majority of the SIDM haloes with $V_\rmn{c,i}(r_\rmn{h})>V_\rmn{h}$ can match the stripping radius criterion, although the median $r_\rmn{max,st}$ at $2\times10^{9}$~$\msun$ is only 10~per~cent larger than $r_\rmn{h}$ due to the high concentration of the haloes prior to core formation.

The constraints of the mass--concentration relation are also evident in the stripped mass fraction. In the bottom-left panel of Fig.~\ref{fig:z0sam} we plot the ratio of mass within $r_\rmn{h}$ post-stripping to the mass with the same radius prior to stripping for subhaloes that satisfy the $r_\rmn{max,st}>r_\rmn{h}$ criterion. Remarkably, all three models require at least 55~per~cent of the mass to be removed from within $r_\rmn{h}$ to fit Crater~II, and specifically 70~per~cent for all three models at $6\times10^{9}$~$\msun$. This finding is qualitatively in agreement with the severe stripping/borderline-disruption argument of \citet{JiA21}. 

Finally, we consider the condensation redshifts (see eqn.~\ref{eqn:zcond} for Crater~II matches in the three models. Most of the SIDM haloes with $M_\rmn{200,i}<10^{10}$~$\msun$ prefer $z_\rmn{cond}<10$. The WDM haloes cluster around the stellar population age measured for Crater~II by \citet{Torrealba16}. We note that pristine gas clouds have been detected at this epoch \citep{Fumagalli2011}, which may provide evidence in favour of such late collapse times. We have therefore developed a picture in which a halo can form very late and be strongly stripped to match Crater~II, especially in WDM. It remains a puzzle as to how likely such a scenario is, although supernova feedback has been argued to be more effective in WDM haloes than in CDM \citep{Orkney22}, and SIDM models that feature a cutoff in the linear power spectrum, through dark acoustic oscillations, in addition to late-time cores -- in effect combining cores with a WDM-style mass--concentration relation -- may provide an additional explanation \citep{Vogelsberger16}.      

In conclusion, we have shown that our stripping model predicts very different requirements to match the various classes of satellites. It is possible to strip most subhaloes to match most satellite galaxies; however, the degree of stripping required is frequently higher than the amount of stripping that can be achieved by the tidal field at the first pericentre, especially for subhaloes of infall $M\rmn{200,i}>10^{9.5}$~$\msun$. The tension is ameliorated to a degree by the switch from CDM to WDM, in which case the lower concentrations at fixed mass allow for greater stripping on a given orbit. The highest density satellites prefer CDM, with some tolerance for WDM subhaloes; they are incompatible with cored subhaloes and thus require gravothermal collapse of low-mass haloes in the SIDM model. Lower density satellites instead have a preference for cores, although for large satellites the cores are not large enough to prevent the need to remove over 50~per~cent of the dark matter from within the observed galaxy.

\section{Matching the MW satellite mass function}
\label{sec:satpop}

We showed in the previous section how the matching potential between observed satellites and infalling subhaloes is very different between dark matter models. In this section we proceed to place these constraints in the context of the expected population of satellite galaxies. We ask whether the distribution of accreted subhaloes' masses and density profiles can match the distribution of the present-day-observed MW satellite mass function, defined from the value of the satellite $M_\rmn{200,i}$ at infall. As a result of this process, we will also be able to derive further constraints on each satellite's properties, given each MW-analogue system will only host so many satellites of a given mass and orbit configuration.

In order to obtain a perfect match to the MW satellite mass function, one would require a deep understanding of the formation history of the MW galaxy, its host halo, and the accretion times of all its satellites. This degree of precision is well beyond a study of this kind, in which none of the host systems can be expected to be a sufficiently accurate representation of the MW system. We will instead develop a simple approach informed by two broad assumptions: (i) the satellites used in this paper are biased towards the most massive infalling subhaloes, and (ii) there is a correlation between halo infall mass and present-day stellar mass, and present the results below. 

\subsection{Populating the halo mass function}

The basic principle of our algorithm is to match as many satellites as possible using the most massive subhaloes available. In practice, we will present a set of 39 observed satellites to the simulation data, and assume that the model combinations that match the largest fraction of the 39 satellites are the best available model prediction.

There are several requirements that we expect for an MW satellite mass function. First, we expect to match as many satellites as possible. Secondly, the hosts should be as massive as possible at infall given that the probability a subhalo hosts a satellite will correlate with infall mass. Below, we present our algorithm for inferring satellite mass functions that match these expectations.

For each of the 102 host systems that we identify in the COCO simulations (see Section~\ref{subsec:halosel}), we apply the accretion time-merger and disc disruption criteria, removing objects that fail these cuts from our sample. We then select one of the 39 satellites at random, assign it to the most massive subhalo that matches the stellar mass--halo mass, $r_\rmn{max,st}>r_\rmn{h}$, and tidal radii criteria for that satellite, and continue to draw satellites at random to assign to the most massive remaining, viable subhalo match until either all satellites have been assigned or there are no further viable matches available. We repeat this process 10 000 times for each host. From these 10 000 realizations we select the iterations that have the joint highest number of successful matches, and from this set of iterations choose the one with the strongest stellar mass--halo mass correlation to be the sole halo mass function for that host going forward. 

\subsection{Inferred mass function properties}

\begin{figure*}
    \centering
    \includegraphics[scale=0.6]{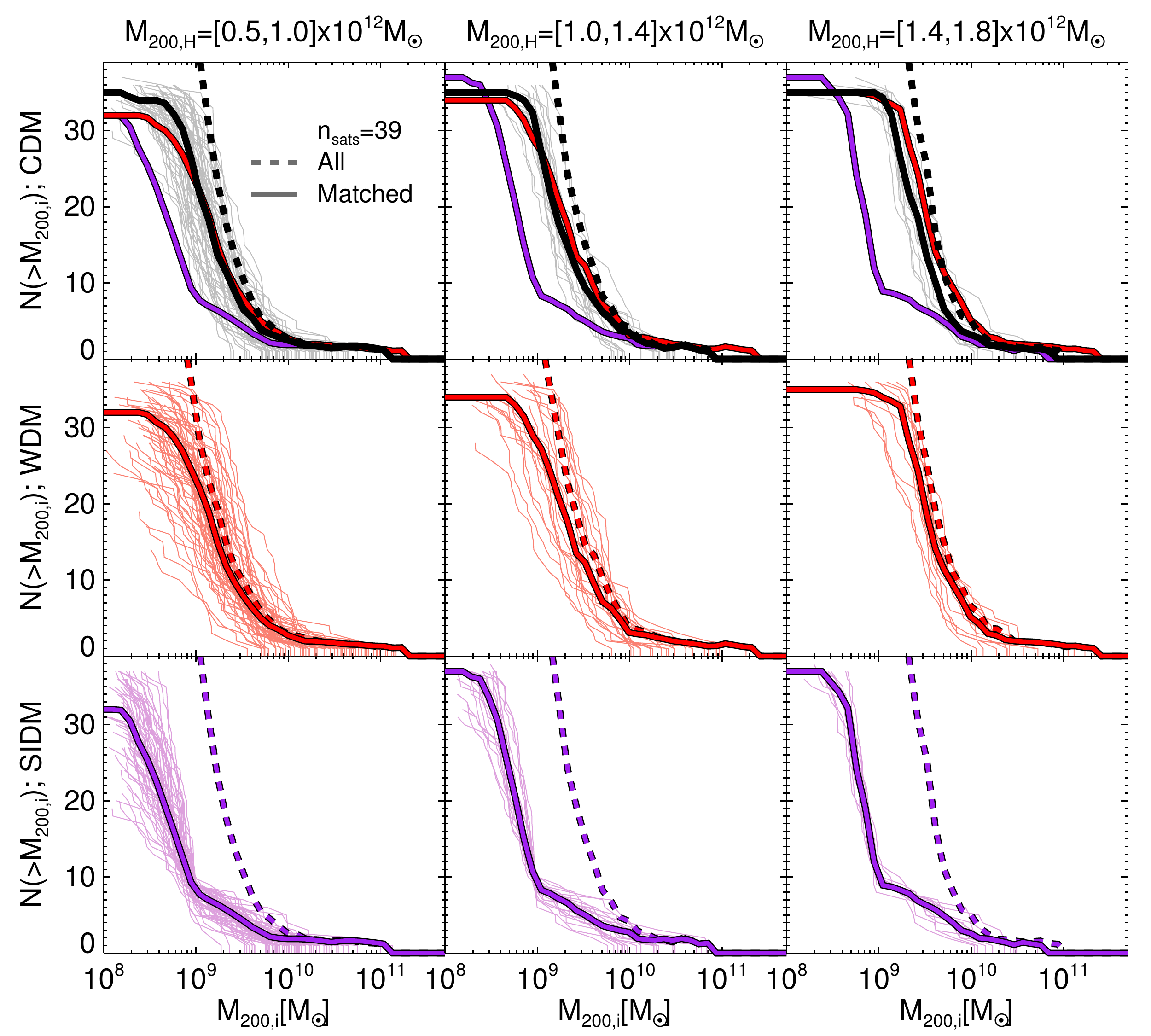} 
    \caption{Infall satellite $M_{200}$ functions inferred for the three models in a series of three bins in host halo mass. The three rows denote data by model: CDM (top), WDM (middle), and SIDM (bottom). Each column denotes a host halo mass bin: [0.5,1]$\times10^{12}$~$\msun$ (left), [1.0,1.4]$\times10^{12}$~$\msun$ (middle), [1.4,1.8]$\times10^{12}$~$\msun$ (right). Individual mass functions are shown as thin lines, and the median mass function as a thick solid line. The thick dashed lines denote the infall $M_{200}$ of all subhaloes, including those that are not assigned a galaxy under this scheme. Finally, the WDM and SIDM median lines are replicated in the top row panels to ease comparison with CDM. The mass functions are truncated from where it is no longer possible to match more satellites for a given host system.}
    \label{fig:mffit}
\end{figure*}

By means of the above algorithm we generate 102 satellite mass functions per dark matter model. We separate these mass functions into three bins in host mass $M_\rmn{200,H}$: $[0.5,1.0]\times10^{12}$~$\msun$, $[1.0,1.4]\times10^{12}$~$\msun$, and $[1.4,1.8]\times10^{12}$~$\msun$. We present the results in Fig.~\ref{fig:mffit}. 

 The CDM and WDM models behave in a similar manner to one another, consistently matching an average of 30--36 satellites out of 39. However, there is a clear difference between the two models in all three mass bins that the WDM curves are shifted to higher subhalo masses than the CDM for $M_\rmn{200,i}>10^{9}$~$\msun$. For example, in the $M_\rmn{200,H}=[1.0,1.4]\times10^{12}$~$\msun$ bin, WDM fits on average 16 satellites with haloes of $M_\rmn{200,i}>10^{9.5}$~$\msun$ compared to 11 for CDM, even though CDM possesses more haloes in total. There is also a smaller gap between median satellite mass function and the {\it total} mass function, as given by the dashed lines, for WDM than for CDM. In the $[1.0,1.4]\times10^{12}$~$\msun$ bin WDM fills on average almost all of the $>5\times10^{9}$~$\msun$ subhaloes, whereas CDM misses 4 on average. The gap between the models is smaller for the $[0.5,1.0]\times10^{12}$~$\msun$ host mass bin, although the gap between the number of matched haloes and the number of matched haloes is still bigger in CDM than in WDM; for the highest mass bin the various discrepancies are similar in size to those at $[1.0,1.4]\times10^{12}$~$\msun$. The crucial challenge for both models in the future will be in the velocity dispersions measured for further satellites: if they require large masses then WDM may fail to generate enough host subhaloes to keep up, whereas if they require small masses CDM will have a more severe too-big-to-fail problem. 
 
 The SIDM model behaves differently with an average of eight satellites per host fit by cored, $M_\rmn{200,i}>10^{9}$~$\msun$ haloes and a further 30 fit by lower mass, collapsed haloes. This leaves an average of 23 unfilled haloes with masses $>10^{9}$~$\msun$. Some of these would likely be filled by adopting a more refined collapse criterion for haloes with masses in the range $[10^{9.0},10^{9.5}]$~$\msun$, as part of a more careful treatment of the mass--collapsed fraction relation; however, the scale of the discrepancy hints that the eventual solution would also require a more extreme cross-section to effect collapse in more massive haloes while preserving the benefits of cored haloes for Crater~II and super-CDM cusps for Willman~I.

We have shown evidence thus far that CDM and WDM have similar potential to host the MW satellite mass function, although WDM prefers to place satellites in more massive subhaloes. Our next step is to ascertain how well each model matches the properties of the observed MW system in general. We consider two key properties: the number of subhaloes that can cool gas to form stars plus the number for which we can match the internal kinematics. The number of MW satellites is a key constraint on WDM models, with models that fail to produce enough satellites considered ruled out. In particular, two recent papers -- \citet{Newton18} and \citet{Nadler20} -- differed in their estimates of the total number of MW satellites by almost a factor of 2, leading to much stronger constraints on WDM in analyses that adopt the \citet{Nadler20} estimate \citep[e.g.][]{Nadler21} than those that instead assume \citet{Newton18} \citep{Enzi21}. The issue of satellite counts is also complicated by the degeneracy between the dark matter model and the host halo mass \cite{Wang12,Kennedy14}, including whether the host is part of a larger galaxy group. We therefore highlight systems for which the parent FoF group mass $M_\rmn{200,FoF}>1.3M_\rmn{200,H}$.  

\begin{figure*}
    \centering
    \includegraphics[scale=0.35]{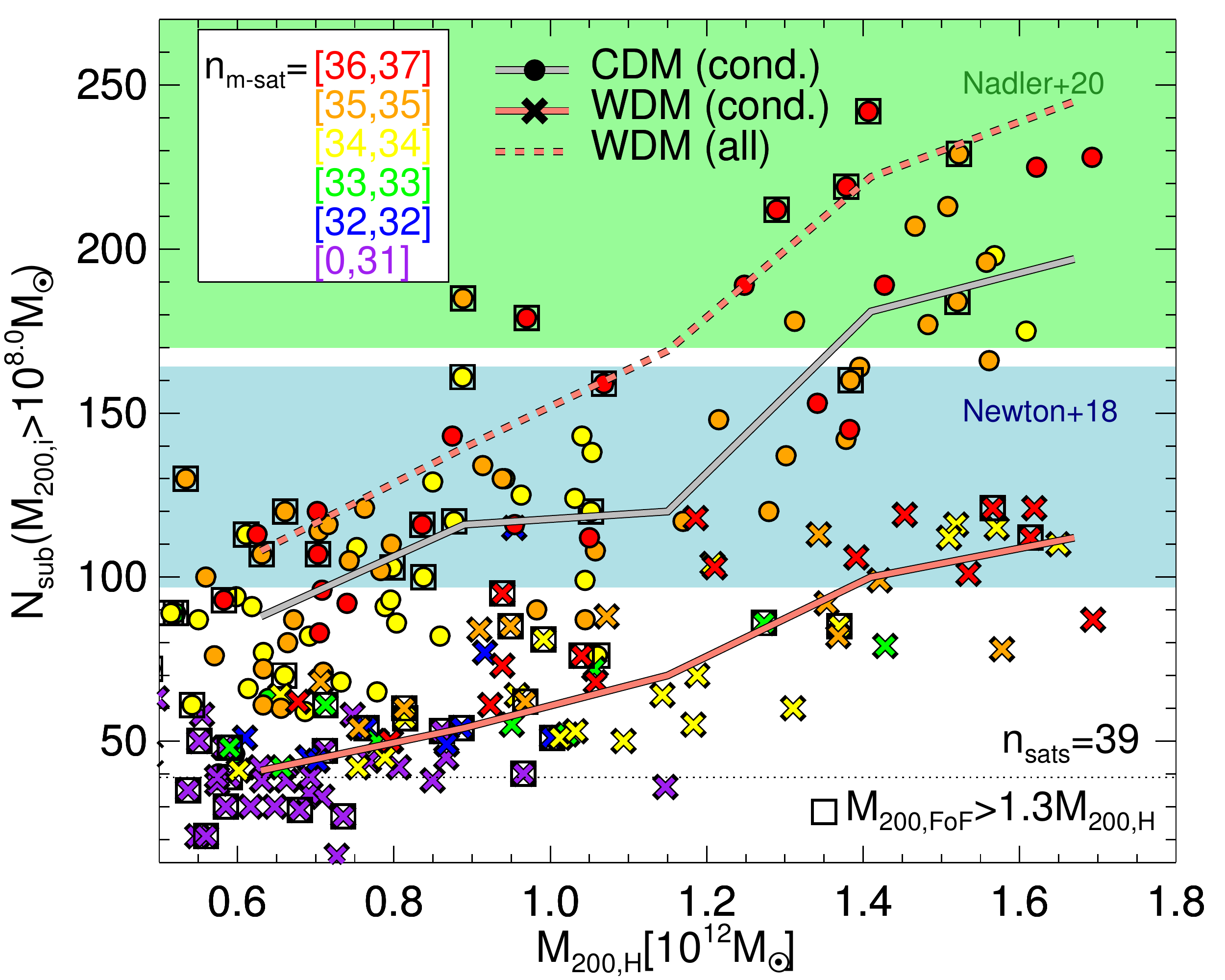}
    \includegraphics[scale=0.35]{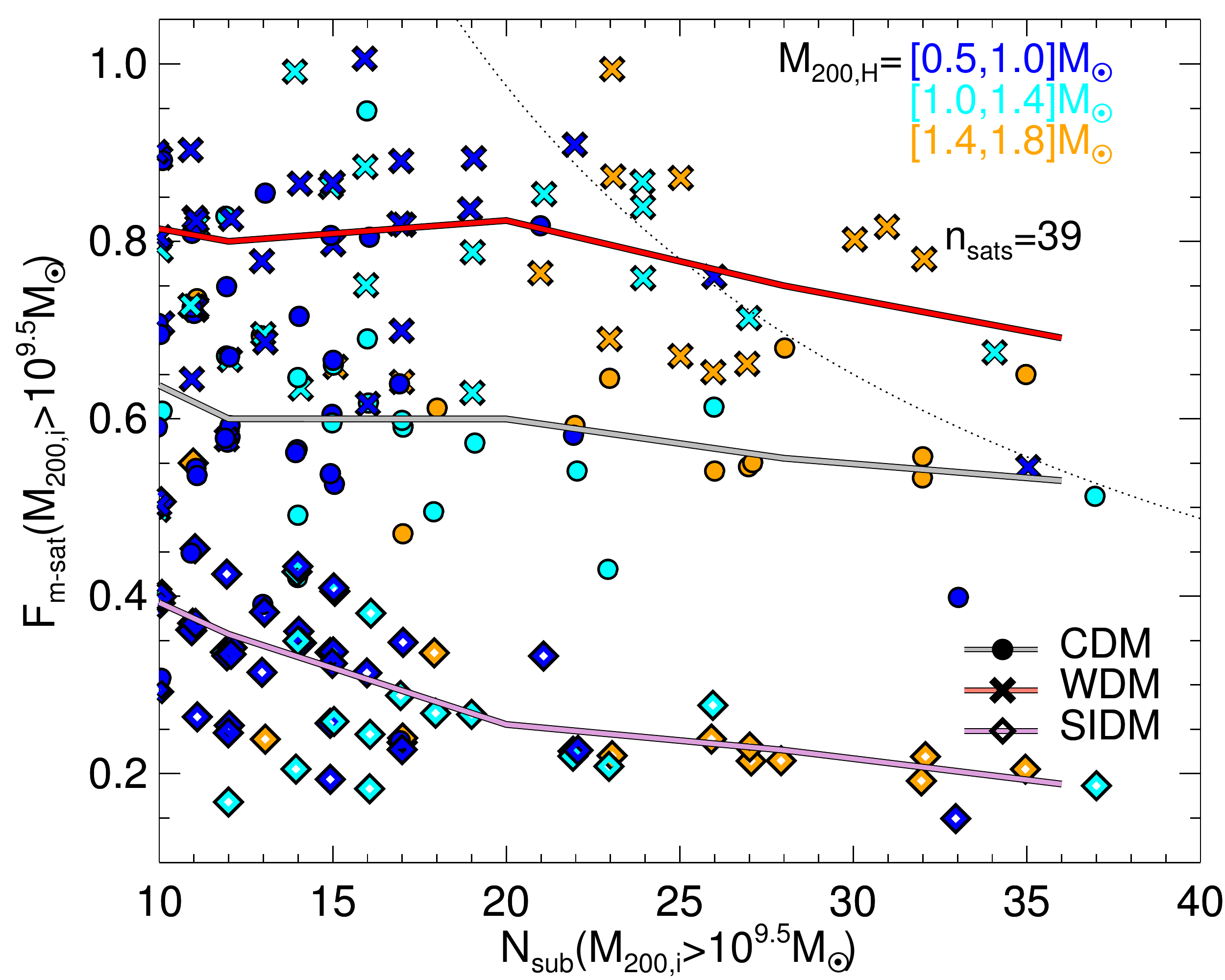}\\

    \caption{Inferred satellite populations for each host halo. {\it Left-hand panel:}  the total number of satellites -- surviving satellites that have achieved baryon condensation -- per host as a function of the host mass when the tidal radius criterion is not applied. CDM hosts data are shown as circles and WDM data are shown as crosses; the symbol colour shows the number of matched satellites as indicated in the figure legends. Hosts for which the parent FoF halo mass, $M_\rmn{200,FoF}>1.3M_\rmn{200,H}$ are enclosed in black squares. The median satellite number -- host mass relations for CDM and WDM are shown as solid grey and light lines respectively. We also include the median number of all surviving WDM subhaloes -- including those for which condensation is not achieved -- as a dashed light red line; the CDM counterpart has a value $>300$ subhaloes for all halo masses and is therefore not shown. We omit the SIDM data from this plot because the number of subhaloes is identical to the CDM value. The blue band is the lower limit on satellite abundance inferred by \citet{Newton18}, and the green band indicates the number of of satellites predicted by \citet{Nadler20}. {\it Right-hand panel}: the fraction of subhaloes with peak $M_\rmn{200,i}>10^{9.5}$~$\msun$ as a function of the number of $M_\rmn{200,i}>10^{9.5}$~$\msun$ haloes. CDM hosts are shown as circles, WDM  hosts as crosses, and SIDM hosts as diamonds. The symbol colour indicates the host mass: blue for $M_\rmn{200,H}<1.0\times10^{12}$~$\msun$, cyan for $1.0\times10^{12}\leq M_\rmn{200,H}<1.0\times10^{12}$~$\msun$, and orange for $M_\rmn{200,H}>1.4\times10^{12}$~$\msun$. The median fraction as a function of halo number is shown with a grey line for CDM, a red line for WDM, and a purple line for SIDM. The solid black line marks $F=N_\rmn{sat}/N_\rmn{sub}$ and the dotted black line denotes $F=0.5\times N_\rmn{sat}/N_\rmn{sub}$.}
    \label{fig:strippop}
\end{figure*}

We therefore compute the total number of subhaloes for each of our 102 systems as a function of host mass and plot the CDM and WDM results in Fig.~\ref{fig:strippop}. The total number of SIDM subhaloes in our analysis is the same as in CDM and is therefore omitted. We note that in SIDM models for which the value of the cross-section at velocities $>150$~$\kms$ -- the characteristic orbital velocity in the MW halo -- is $\gsim1$~cm$^{2}$gr$^{-1}$ \citep{Vogelsberger12,Rocha13}, interactions between the satellite and host dark matter particles can evaporate subhaloes, leading to a suppression of the subhalo mass function \citep{Vogelsberger12}. \cite{Turner21} found that the AqA3-vd100 simulation predicted up to 10~per~cent of satellites evaporate in vd100. We do not take this account into effect, but state that this issue will grow for still more extreme cross-section functions. 

The increase in the number of subhaloes for CDM over WDM allows for many more objects that could match each satellite, without accounting for differences in the internal subhalo kinematics. We anticipate that the satellites discovered to date correspond to the MW's most massive subhaloes, given the expected correlation between luminosity and halo mass, and so our MW-analogue system's massive subhaloes should all host satellites. We therefore compute the fraction of each host's subhaloes of mass $M_\rmn{200,i}>10^{9.5}$~$\msun$ that are matched to satellites as a function of the total number of  $M_\rmn{200,i}>10^{9.5}$~$\msun$, and plot this alongside the subhalo counts in Fig.~\ref{fig:strippop}.      

There is a factor of $\sim2$ difference in the total number of atomic-cooling subhaloes between CDM and WDM. Interestingly, the number of CDM subhaloes in the mass range of the MW -- $[1.0,1.4]\times10^{12}$~$\msun$ \citep{Callingham19} -- marginally exceeds the reported lower limit of the \citet{Nadler20} satellite estimate whereas the WDM count is in general lower than the \citet{Newton18} estimate, although five WDM systems nevertheless do generate enough satellites to agree with \citet{Newton18}. We therefore reassert that the assumptions incorporated into estimates of MW satellite counts are highly non-trivial for WDM constraints. 

Despite the large disparity in the total number of satellite galaxies, the gap between the two models in the number of matched satellites -- as indicated by symbol colour -- is much smaller, at $>32$ matches for both models at $M_{200}>10^{12}$~$\msun$. However, there is a significant trend for CDM to match marginally more satellites than WDM: in the MW mass range $[1.0,1.4]\times10^{12}$~$\msun$, 17 CDM systems match at least 35 out of 39 satellites compared to  9 WDM systems and 19 SIDM systems (SIDM not shown). Some 67~per~cent (60~per~cent) of CDM (WDM) systems with $M_\rmn{200,FoF}>1.3M_\rmn{200,H}$ return more satellites than expected from the median relation across all systems, thus we have some evidence in both models that MW systems in LG environments host more satellites than isolated MW-mass haloes. Finally, we show that uncertainty in our prescription for which haloes can undergo condensation and form stars may permit the number of WDM subhaloes to increase by up to a factor of 3, and therefore a careful treatment of halo trees and collapse times will be required to obtain precise predictions.   

The following question is whether, having succeeded in hosting most of the observed satellites in simulated subhaloes, each satellite has been placed in a halo appropriate to its stellar mass. We have already indicated in Fig.~\ref{fig:mffit} that the WDM model populates the highest mass haloes most frequently, ahead first of CDM and then SIDM. We place this in a quantitative context in the right-hand panel of Fig.~\ref{fig:strippop}, in which we plot the fraction of $>10^{9.5}$~$\msun$ haloes that achieve a match. The WDM model consistently fills 80~per~cent of the available subhaloes in this mass range, compared to 60~per~cent for CDM and 20~per~cent for SIDM, with slightly lower fractions at the highest host halo masses since they possess more subhaloes and thus have a larger number of luminous satellites than is the case for the observed MW system.

Given that all of these haloes with $M_\rmn{200,i}>10^{9.5}$~$\msun$ are expected to be luminous, we would infer that in the CDM 40~per~cent of the satellites in massive subhaloes have yet to be detected compared to 20~per~cent of WDM satellites. The vd100 SIDM model requires a further 80~per~cent to be detected, or from 11 to 20 objects, which could be consistent with massive, diffuse Crater~II and Antlia~II type satellites at large distances.  Alternatively, a successful SIDM model will require a more extreme cross-section than vd100 in order to spur gravothermal collapse in more massive subhaloes. 

In conclusion, we have shown that the CDM model matches the largest number of satellites included in our analysis, likely by virtue of hosting more subhaloes, but WDM has the highest subhalo occupancy fraction at high masses when considering detected satellites. The vd100 model predicts that many massive, cored subhaloes are yet to be discovered, and that the current satellite population is biased towards low-mass haloes. Finally, the agreement in the total number of satellites is very sensitive to the assumptions adopted in the observational estimates.

\section{Satellite property estimates}
\label{sec:satprops}

We have thus far demonstrated the scope that each model has to match each of the 39 satellite galaxies considered in this paper, and expanded that analysis to estimate the satellite mass function. In this section we will use the combination of our results from fitting the circular velocity profiles -- both from satellite masses and from mass function considerations -- and mass function considerations to derive predictions for satellite properties.    

\begin{figure}
    \centering
    \includegraphics[scale=0.55]{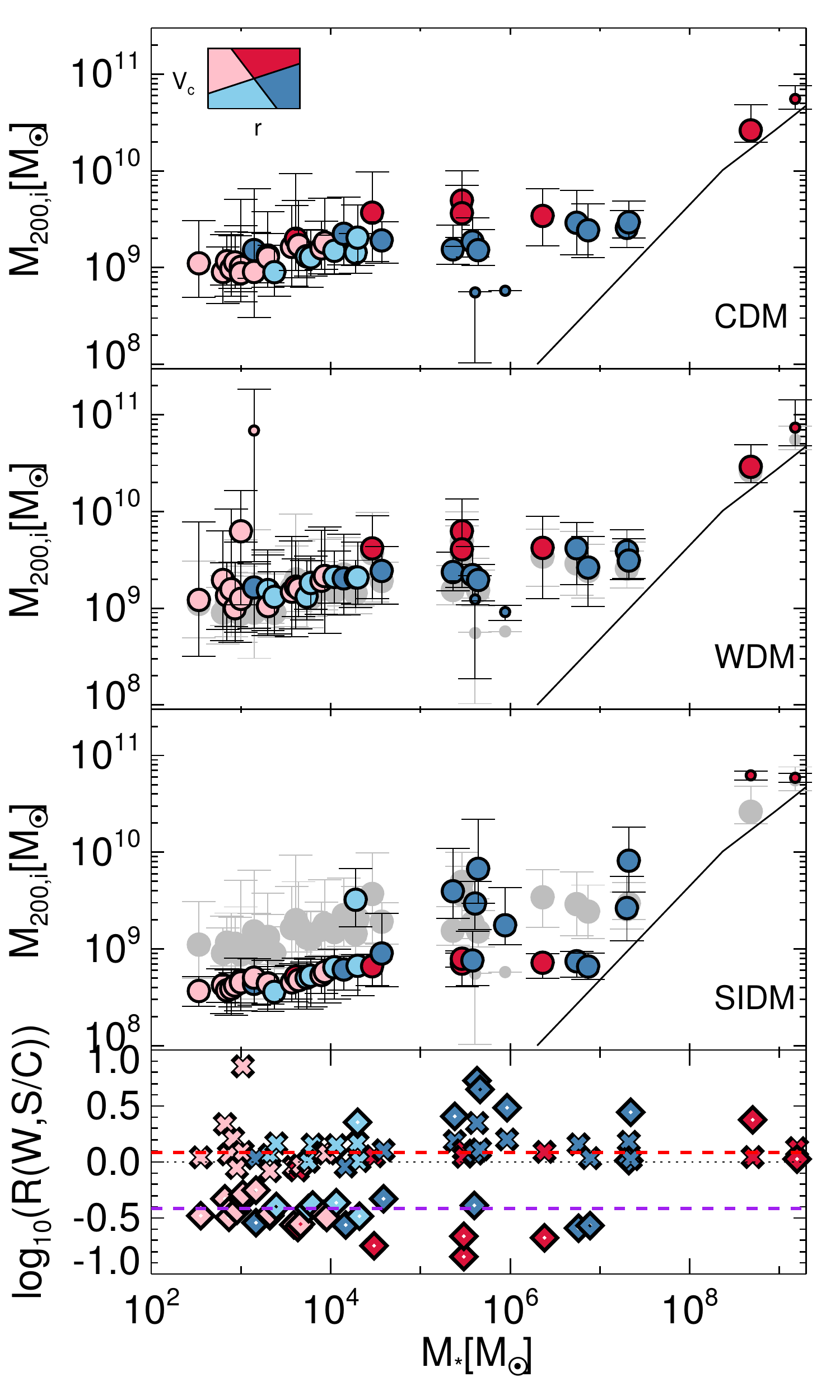} 

    \caption{Estimates for satellite infall mass $M_\rmn{200,i}$ as a function of stellar mass. The top three panels present median $M_\rmn{200,i}$ plus 68~per~cent regions for CDM,  WDM and SIDM from top to bottom; the bottom panel shows the ratio of the median WDM (crosses) and SIDM (diamonds) $M_\rmn{200,i}$ with respect to CDM. The CDM data are reproduced in the WDM and SIDM panels as grey silhouettes, and the median mass ratios for WDM and SIDM across all 39 satellites are shown as dashed red and purple lines respectively in the bottom panel. The symbol colour indicates the portion of $V_\rmn{h}$--$r_\rmn{h}$ space occupied by each satellite via the inset in the top panel: high-density-large size satellites are shown in red, high-density-small-size in pink, low-density-large-size in dark blue, and low-density-small-size in light blue. The solid black lines in the upper three panels indicate the minimum $M_\rmn{200,i}$ permitted for $M_{*}$ due to star formation efficiency constraints. Satellites for which there are fewer than 25 matches are shown as small symbols, and the remainder are shown as large symbols.}
    \label{fig:stripM200}
\end{figure}

We begin with predictions for the infall mass, $M_\rmn{200,i}$. For a given satellite, we compute the median value of $M_\rmn{200,i}$ across all of the host systems that possess an analogue of that satellite, for up to 102 systems; we also compute the 68~per~cent range. We plot the results as a function of the measured stellar mass of each satellite, $M_{*}$, in Fig.~\ref{fig:stripM200}. Throughout this section we assume that the halo profiles are not affected by the condensation of gas into stars.

In this framework, the CDM model expects that the median infall mass of almost all satellites is $>10^{9}$~$\msun$; note that the inclusion of more observed satellites as kinematic data become available will lower the $M\rmn{200,i}$ estimate for the fainter satellites. The two glaring exceptions are Crater~II and Antlia~II at $M_{*}\sim10^{5.5}$~$\msun$, which both expect to have masses below $10^{9}$~$\msun$ even in the 68~per~cent ranges. Our results are therefore broadly consistent with \citet{Hayashi22}, who found that the central density of these two satellites is significantly lower than for much fainter dSphs. In both cases we find fewer than 25 matches across our 102 systems, adding to their peculiarity.

Among the remainder of the satellites, there is a significant correlation between infall mass and stellar mass, which is at least partly by construction given the abundance matching piece of our luminosity function. At present this relation spans the range $[1,3]\times10^{9}$~$\msun$ from low stellar mass to the brightest dSph, and in our algorithm will become steeper with the inclusion of faint satellites. Denser satellites (pink and red) unsurprisingly anticipate more massive haloes than then less dense counterparts at fixed brightness. The estimates for the LMC and the Small Magellanic Cloud (SMC) are $[4,10]\times10^{10}$~$\msun$ and $[2,6]\times10^{10}$~$\msun$ respectively. However, these satellites are more likely to have had their profiles changed by baryonic processes resulting from galaxy formation not included in our analysis, and should therefore be treated with caution.

The WDM results are qualitatively the same as for CDM, from the approximate luminosity--halo mass slope to behaviour of low-density versus high-density satellites. The one significant  difference is that each dSph is on average 25~per~cent more massive in WDM than for CDM, which is expected due to the combination of its mass--concentration relation convolved with the orbit distribution and the mass function requirements. WDM obtains only marginally more Crater~II analogues than CDM, but its $M_\rmn{200,i}\sim10^{9}$~$\msun$ so there is a degree of improvement. SIDM instead returns very different results to CDM, with many large-size-low-density satellites hosted in massive cored subhaloes and the remainder confined to gravothermally collapsed haloes that are on average less than half the mass of their CDM counterparts. \citet{Correa21} instead found all classical satellites to have undergone a degree of gravothermal collapse, possibly due to a marginally more extreme self-interaction cross-section; we agree that many classical satellites would require $M_\rmn{200,i}<10^{9}$~$\msun$.

\begin{figure}
    \centering
    \includegraphics[scale=0.55]{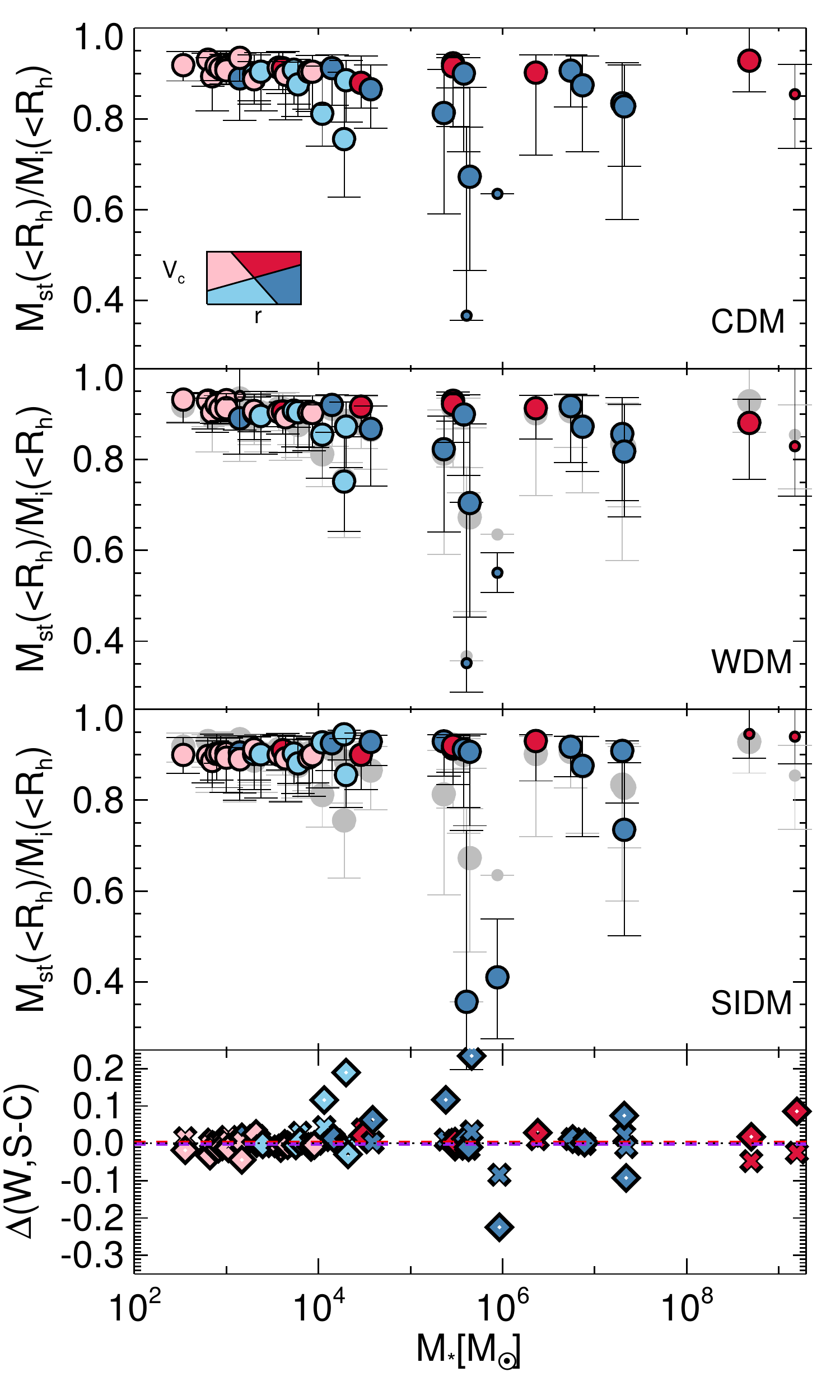} 

    \caption{Estimates for satellite mass ratios within $r_\rmn{h}$ as measured between the stripped time and the infall time. The plot format is the same as for Fig.~\ref{fig:stripM200} except that in the bottom panel we present the median difference in the stripped ratios between CDM and the other two models.}
    \label{fig:stripMRat}
\end{figure}

We next turn to the stripping rates of our satellites.  We compute the ratio of the mass within $r_\rmn{h}$ post-stripping and at the infall time -- $M_\rmn{st}(<r_\rmn{h})/M_\rmn{i}(<r_\rmn{h})$ -- and repeat the process applied for $M_\rmn{200,i}$ in Fig.~\ref{fig:stripM200} with this quantity. We present the results in Fig.~\ref{fig:stripMRat}. The one difference from Fig.~\ref{fig:stripM200} is that we compare CDM to the other two models by computing the difference between the median values rather than the ratio. 

The majority of CDM mass ratio estimates are above 80~per~cent, thus only 20~per~cent of the mass has been lost; the median stripped mass is $\sim25$~per~cent. There is evidence that lower density satellites have experienced more stripping than higher density satellites, with a median stripping rate of 15~per~cent for the former and 10~per~cent for the latter. Crater~II is once again an outlier, with a median 30~per~cent stripping rate and some instances where more than 60~per~cent of this mass has been lost. Antlia~II instead expects to have lost no more than 35~per~cent of its mass for this halo selection. 

The WDM predictions are remarkably similar to CDM, with a median difference across all satellites of only 1~per~cent. The higher masses and lower concentrations of the WDM fits cancel out the need for more or less stripping relative to CDM. SIDM shows a bigger difference from CDM, with less stripping for most satellites first due to the need to increase the concentration to compensate for the much lower halo masses in faint satellites, and second because large bright satellites are fit with cores and so do not need stripping. The exceptions are again Crater~II and Antlia~II, which despite their cores, typically lose half of their mass with $r_\rmn{h}$. We recognize this phenomenon as an expression of the halo stripping problem introduced by \citet{Errani22} and \citet{Borukhovetskaya22}, namely that $r_\rmn{h}$ is larger than the $r_\rmn{max}$ of most viable hosts and is therefore larger than the core radius we impose. In other words, the amount of stripping required would significantly affect its stellar component, causing it to be less spatially extended than that required by observations.  We therefore show that all three models have difficulty fitting these two satellites, although coupling cores with the \citet{JiA21} disruption result is a viable option.    

\begin{figure}
    \centering
    \includegraphics[scale=0.55]{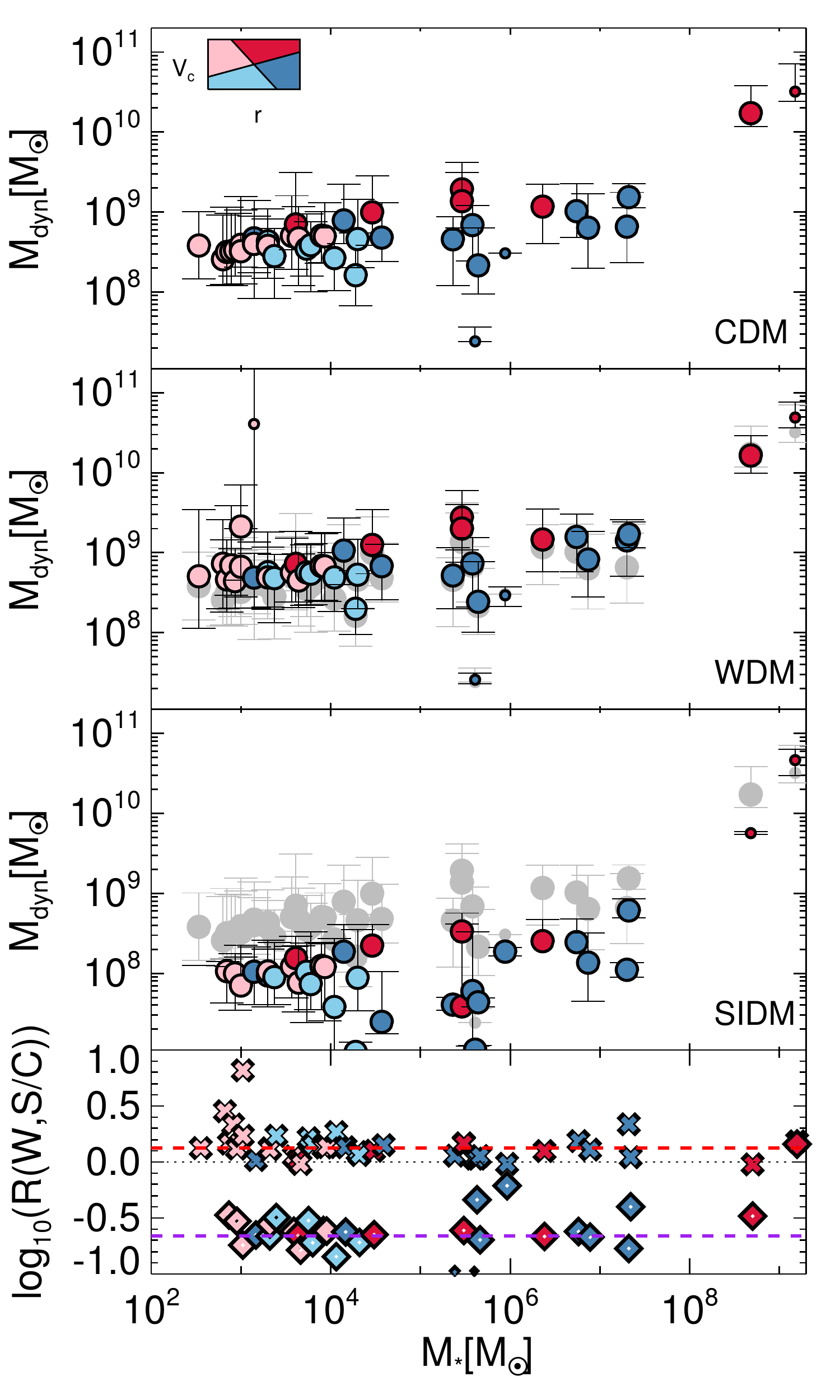} 

    \caption{Estimates for satellite present-day dynamical mass, $M_\rmn{dyn}$ as a function of stellar mass. The plot format is the same as for Fig.~\ref{fig:stripM200}.}
    \label{fig:stripMdyn}
\end{figure}

Thus far in this paper we have focused on the mass measured at infall, $M_\rmn{200,i}$. By contrast, the redshift zero mass is relevant for analysing the present day kinematics of the satellite. This present day mass can be difficult to define due to defining a boundary between the subhalo and the host halo. We choose to define a $z=0$ dynamical mass, $M_\rmn{dyn}$, as the total mass enclosed within the satellite tidal radius. We repeat the process applied for $M_\rmn{200,i}$ in Fig.~\ref{fig:stripM200} and present the results in Fig.~\ref{fig:stripMdyn}.

The relationship between the satellites $M_\rmn{dyn}$ is broadly the same as that of  $M_\rmn{200,i}$ except that stripping has reduced the masses by a factor of $\sim5$. The one exception is Antlia~II, for which the halo mass-loss is less, at around a factor of 2. WDM again follows a similar trend except that the haloes are more massive than in CDM: the average difference is 30~per~cent and in some cases is a full factor of two higher. SIDM, by contrast, requires all of the satellites to be less massive than $10^{9}$~$\msun$, regardless of whether they exhibit cores or gravothermal-collapse cusps. A vd100 SIDM halo of $M\rmn{200,i}>10^{9}$~$\msun$ would have $r_\rmn{h}=[1-3]$~kpc  and $V_\rmn{h}=[20,30]$~$\kms$, and remarkably there are no observed satellites in our sample that have this pair of parameters (see Fig.~\ref{fig:satsum}). We therefore find that, paradoxically, vd100 SIDM suffers from a severe too-big-to-fail problem once mass-luminosity matching is taken into account . 

\begin{figure}
    \centering
    \includegraphics[scale=0.55]{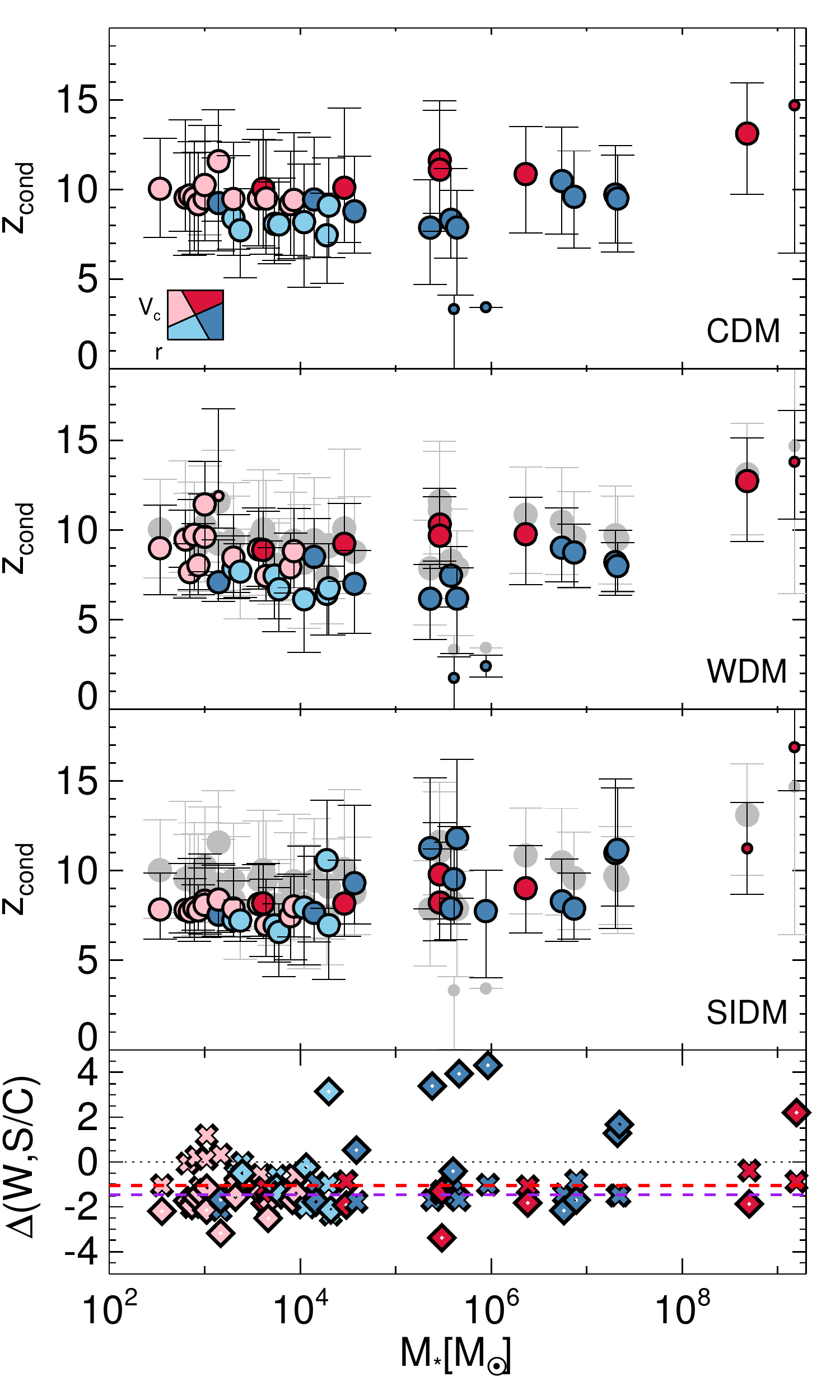} 
    \caption{Estimates for satellite condensation redshift $z_\rmn{cond}$ as a function of stellar mass. The plot format is the same as for Fig.~\ref{fig:stripM200} except that in the bottom panel we present the median difference in the stripped ratios between CDM and the other two models.}
    \label{fig:stripZCond}
\end{figure}

We end this section with a presentation of the condensation redshift, $z_\rmn{cond}$. It has been shown that WDM haloes form later than their CDM counterparts \citep{Lovell12,Lovell17b,Maccio19}, which may in principle be detectable from studies of stellar populations. We compute the distributions of $z_\rmn{cond}$ for the 39 satellites in our three models and present the results in Fig.~\ref{fig:stripZCond}. 

The majority of the CDM satellites undergo condensation well before reionization at $z=6$, and typically in the range $z_\rmn{cond}=[8,12]$. There is very little trend with stellar mass. Less dense satellites form marginally later than their overdense counterparts, which tracks the lower densities of later forming haloes. Finally, Crater~II and Antlia~II are strong outliers, appearing to only start forming stars at redshifts $z_\rmn{cond}<5$ and therefore needing to retain gas well beyond reionization. 

The WDM satellites start to form later than CDM as expected, typically by 1.5 units in redshift to $z_\rmn{cond}=[7,11]$. Antlia~II and Crater~II remain anomalous, although marginally less so than in CDM due to the later condensation redshifts of the other satellites. SIDM satellites instead show a bimodal distribution, with low-density, cored objects first condensing before $z=10$ and gravothermally collapsed objects delayed until after $z=9$ on average due to their small masses, albeit with much larger scatter than CDM or WDM. Interestingly, the application of SIDM cores and the WDM power spectrum cutoff drive the condensation redshift of low-density satellites in opposite directions relative to CDM. Both models require subhaloes more massive than in CDM, but SIDM requires high-density satellites with small scale radii in order not to undershoot the satellite density while WDM instead requires haloes of larger scale radii which translates into later formation times. The time-scales involved are of order 100~Myr and therefore are very difficult to measure with current techniques; nevertheless, it is interesting to have a potential discriminator between these two models as a target for future work.  

\section{Conclusions}
\label{sec:conc}

The properties of the MW satellites remain one of the most important tests for discriminating between dark matter models. Three of the most popular dark matter models -- CDM, WDM, and SIDM -- each make separate predictions for the satellite density profiles: concentrated cusps for CDM, less concentrated cusps for WDM, and cores plus gravothermally collapsed cusps for SIDM models with cross-sections $\gsim50\rmn{cm^2g^{-1}}$. The question posed to each model is whether its selection of density profiles can match the full range of MW satellites, and in the correct abundance to match the MW satellite circular velocity function. 

We approach this problem in the context of these three dark matter models through the COCO simulations. The first simulation adopted CDM, and the second simulation used a WDM thermal relic with a mass of 3.3~keV; this simulation is a good approximation to structure formation in the sterile neutrino cosmology for which the sterile neutrino mass is 7.1~keV and the lepton asymmetry $L_6=10$ \citep{Lovell20}. We also develop an SIDM data set that approximates the vd100 model used in \citet{Zavala19a} and \citet{Turner21}.  For this model we convolve their simulation results with the COCO-CDM simulation, determining both cored profiles and gravothermally collapsed cusps. For these three models we identify 102 MW analogue host haloes and the subhaloes of maximum mass $>10^{8}$~$\msun$ whose descendants were located within 300~kpc of these hosts at $z=0$. We approximate these subhaloes prior to infall with fitted NFW profiles, plus in the case of SIDM a method is used to assign a cored or gravothermally collapsed profile, and then stripped these profiles to match the masses of 39 observed MW satellites using the method of \citet{Errani21}. We also perform a stripping analysis of a set of isolated haloes selected at $z=1$. We identified viable host subhaloes for satellites as those for which the stripped $r_\rmn{max,st}$ is greater than the observed satellite radius, $r_\rmn{h}$, that were sufficiently massive to enable cooling, and that the post-stripping tidal radii were commensurate with the subhaloes' orbits.  
 
The WDM and CDM subhaloes that accrete onto MW host-analogues are able to fit the $r_\rmn{max,st}>r_\rmn{h}$ criterion of most satellites well, with WDM preferring higher mass subhaloes to compensate for their lower concentrations. Where they differ is in whether the subhalo orbit tidal field is strong enough to effect the required stripping: the lower concentrations of WDM objects enable the satellite $V_\rmn{h}$ to be matched with less stripping, and so at fixed halo mass and orbit the WDM model is able to match more satellites. Both models fail to match the massive underdense satellites, Crater~II and Antlia~II, except for the unlikely case that their pre-infall masses are $<10^{9}$~$\msun$.

Cored SIDM haloes are much more readily able to match these specific, low-density satellites, but the cores are still sufficiently small that at least 60~per~cent of the mass within the half-light radius has to be stripped away. The price paid to fit the underdense, massive satellites in SIDM is that dense satellites cannot be fit with cored, massive subhaloes and therefore rely on gravothermal collapse, which in the vd100 model only occurs for subhaloes of infall mass $<10^{9}$~$\msun$. Finally, we repeat this process with $M_{200}>10^9$~$\msun$ subhaloes identified at $z=1$ and find that 5~per~cent of these WDM haloes could indeed fit Crater~II, albeit while being stripped of over 50~per~cent of their half-light radius mass. Part of the discrepancy could be that the MW host-accreted subhaloes formed earlier and had higher densities. We also note that these Crater~II analogues had the same, late ($z\sim3$) cooling times as measured for the Crater~II stellar population reported by \citet{Torrealba16}.

We then proceed to consider the infall mass function in the three models. We develop an algorithm to insert viable satellite galaxies into the subhalo mass function by assuming that the observed satellites occupy the most massive satellites at infall, and that there is a correlation between infall mass and present stellar mass. We find that WDM MW-mass hosts typically fit 34 of the 39 satellites we considered, compared to 35/39 satellites in CDM and 37 in SIDM, but did so in very different ways. The SIDM model placed many satellites into $<10^{9}$~$\msun$ gravothermally collapsed haloes and so left high-mass haloes vacant. WDM could fill 80~per~cent of the haloes of mass $>10^{9.5}$~$\msun$ with satellites using this algorithm. This is compared to 60~per~cent for CDM, which instead hosts a larger proportion of satellites in lower mass haloes. CDM therefore requires the discovery of 8-12 more satellites in this mass range compared to 4-6 for WDM and 14-28 for SIDM. We also compute estimates for the total number of MW satellites in the CDM and WDM models, and show that the difference between the two is similar to the current variation in the observationally inferred satellites counts.

Finally, we produce estimates for four properties of the satellites in the three cosmologies: infall mass, post-stripping mass-loss within $r_\rmn{h}$, post stripping dynamical mass, and the condensation redshift, which is the first redshift at which the halo can form stars. Both infall masses and post-stripping dynamical masses are on average 25~per~cent higher for WDM than CDM. The infall masses of massive, low-density satellites are high in cored SIDM haloes but suppressed below $10^{9}$~$\msun$ when gravothermal collapse is required, the difference between the two is erased by subsequent stripping of the halo outskirts. We therefore argue that, under the SIDM model, one of at least three conditions should be met: (i) there exists an undetected population of large ($\sim2$~kpc), cored satellites that are not stripped, (ii) all massive satellites were accreted at sufficiently early times to have undergone significant stripping, or (iii) the distribution  of satellite densities reflects a fine tuning of the self-interaction cross-section velocity dependence \citep{Meshveliani22}.

The stripping rate within $r_\rmn{h}$ differs much less between models, although the Crater~II and Antlia~II satellites in SIDM are still required to lose substantially more of their mass than other satellites and so even in SIDM remain anomalous, unless they are on the cusp of disruption \citep{JiA21}. Finally, the condensation redshift of the satellites is delayed systematically in WDM relative to CDM. SIDM gravothermal collapse haloes are delayed still further, but non-collapsed haloes form earlier than in CDM. Therefore, SIDM cores require galaxies be older than in CDM whereas WDM cusps require they be younger.    

Our primary conclusions from this work are that SIDM cores offer the easiest  dark matter-physics resolution to the question of fitting underdense, bright satellites at the expenses of the gravothermal collapse and a lack of large, low-density galaxies, and that the WDM model is the most efficient at hosting bright satellites in massive subhaloes due to the relationship between orbit size and the required stripping rate. It is unclear whether the WDM power spectrum cutoff alone is sufficient to match the Crater~II velocity dispersion, but does make it easier for subsequent supernova feedback to lower the density further than is the case for CDM \citep{Orkney22}; we defer further investigation of this possibility to future work. The SIDM cosmology may be more viable in models where the velocity-dependent cross-section is higher than in vd100 in order to facilitate gravothermal collapse in more massive subhaloes. 

The next key developments in the computational methodology will be to finesse the stripping method of \citet{Errani21} for cored profiles and to finesse the computation of the orbits, both to account for more precise calculations of the first pericentre to follow satellites around subsequent pericentres and to employ a more detailed model of gravothermal collapse. The most crucial results will be from future searches for satellites: large numbers of dense satellites will prefer gravothermal collapse SIDM, and the absence of large numbers of bright satellites will instead favour WDM.  

\section*{Acknowledgements}

We would like to thank Rafael Errani, Victor Forouhar-Moreno and Carlos Frenk for valuable suggestions and comments. MRL acknowledges support by a Grant of Excellence from the Icelandic Research Fund (grant number 206930).

\section*{Data Availability}

The COCO-CDM and COCO-WDM simulations were originally published in  \citet{Hellwing16} and \cite{Bose16a}. Requests for access should contact Wojtek Hellwing. The data from the Aq-A3-vd100 simulation can be shared on reasonable request from M.~R.~Lovell.



\bibliographystyle{mnras}








\bsp	
\label{lastpage}
\end{document}